\documentclass[reprint,aps,prl,nofootinbib,superscriptaddress,floatfix]{revtex4-2}

\usepackage[utf8]{inputenc}
\usepackage[T1]{fontenc}
\usepackage{amsmath,amssymb}
\usepackage{xcolor}
\usepackage{graphicx}
\usepackage{appendix}
\usepackage{adjustbox}
\usepackage{bm}
\usepackage{epstopdf} 
\usepackage[colorlinks=true, allcolors=blue]{hyperref}
\usepackage{orcidlink}
\usepackage{multirow}
\usepackage{booktabs}
\hypersetup{
  pdftitle={A Predictive Nuclear Effective Field Theory from Wilsonian Renormalization Group Principles I},
  pdfauthor={Chen-Can Wang, Jia-Ai Shi, and Bing-Nan Lu}
}

\newcommand{\nc}{\newcommand}       % new command
\nc{\no}[1]{:\!\! #1 \!\!:}

\newcommand{\gscas}{\affiliation{Graduate School of China Academy of Engineering Physics, Beijing 100193, China}}

\begin{document}

\title{A Predictive Nuclear Effective Field Theory \\
from Wilsonian Renormalization Group Principles I: \\
Medium-mass nuclei and nuclear matter}

\author{Chen-Can Wang}
 \gscas
\author{Jia-Ai Shi}
 \gscas
\author{Bing-Nan Lu}
\email{bnlv@gscaep.ac.cn}
 \gscas

\date{\today}

\begin{abstract}
The predictive power of nuclear effective field theory (EFT) from few- to many-body systems is substantially restricted by regulator dependence. Wilsonian momentum-shell renormalization group (RG) offers a natural framework for cutoff-independent predictions, but its many-body implementation requires regulating single-particle momentum modes, which breaks Galilean invariance. RG scaling analysis shows that this symmetry breaking is encoded in irrelevant operators whose effects can be included explicitly or absorbed into higher-body interactions. Guided by Wilsonian RG, we construct a regularization scheme for chiral EFT contact interactions and predict binding energies of nuclei up to $^{40}$Ca using interactions calibrated only to $A\leq3$ systems. The predicted energies vary at the percent level and agree with experiment for $300\leq\Lambda\leq450$~MeV. The same scheme reduces the overbinding of symmetric nuclear matter obtained with conventional relative-momentum regulators. Our results suggest that some model dependence observed in nonrelativistic many-body calculations originates from treating EFT vertices as static, Galilean-invariant potentials rather than as field operators regulated consistently across particle-number sectors.
\end{abstract}

\maketitle

\textit{Introduction}---Nuclear \textit{ab initio} calculations based on chiral effective field theory (EFT) provide a powerful, systematically improvable description of atomic nuclei and nuclear matter~\cite{Weinberg1990_PLB251-288,Epelbaum2009_RMP81-1773,Navratil2016_PS91-053002,Machleidt2011_PhysRept503-1, Hammer2020_RMP92-025004,Machleidt2023_FBS64-77,Papenbrock2024_arXiv2410-00843}. The theory introduces a regulator cutoff $\Lambda$ to separate resolved from unresolved momentum scales. A key consistency requirement of nuclear EFT is the regulator  independence of observables, namely its renormalizability~\cite{Machleidt2020_EPJA56-95,Furnstahl2021_FBS62-72}. Its formal proof, however, faces significant   challenges arising from the nonperturbative nature of the nuclear many-body systems.

Renormalization studies in nuclear EFT have largely focused on nucleon-nucleon (NN) scattering~\cite{Birse1999_PLB464-169,Epelbaum2017_NPB925-161,Contessi2024_PLB856-138903} and few-body systems~\cite{Bedaque1999_NPA646-444,Bedaque1999_PRL82-463,Bedaque2000_NPA676-357,Platter2004_PRA70-052101,Platter2005_PLB607-254,Kirscher2013_PLB721-335,Epelbaum2017_EPJA53-98,Song2017_PRC96-024002,Bazak2019_PRL122-143001}. Several renormalization prescriptions have been developed, and debate continues over whether the cutoff should be taken far above the EFT breakdown scale $\Lambda_b$ (${\sim}500$~MeV for chiral EFT)~\cite{Nogga2005_PRC72-054006,Long2011_PRC84-057001,Long2012_PRC86-024001,Long2012_PRC85-034002,Yang2025_PRC112-014004} or kept near that scale~\cite{Epelbaum2009_EPJA41-341,Epelbaum2018_EPJA54-186,Epelbaum2020_EPJA56-152}, and whether subleading interactions should be treated perturbatively~\cite{Kaplan1998_PLB424-390,Long2011_PRC84-057001,Long2012_PRC86-024001,Song2017_PRC96-024002} or iterated nonperturbatively~\cite{Kirscher2010_EPJA44-239,Lynn2017_PRC96-054007}. Related studies extending to $A>4$ are scarce, with applications to nuclear matter~\cite{Dyhdalo2016_PRC94-034001,Yang2023_EPJA59-233,Valderrama2026_PRC114-034010} and nuclei such as $^{16}$O up to next-to-leading order (NLO)~\cite{Schiavilla2021_PRC103-054003,Yang2021_PRC103-054304,Sanchez2023_PRC107-064307,Contessi2025_arXiv2505-09299}. The general conditions for cutoff-independent many-body predictions, however, are not yet established.

Besides formal studies, a practical concern in modern nuclear \textit{ab initio} calculations is the invariance of predictions under finite cutoff variations, as required by renormalization group (RG) invariance. Such calculations typically combine nonperturbative many-body methods with high-precision chiral NN interactions at N$^3$LO and beyond, regulated at relatively low cutoffs around $300$--$500$~MeV~\cite{Entem2003_PRC68-041001,Epelbaum2005_NPA747-362,Epelbaum2015_EPJA51-53,Entem2017_PRC96-024004,Reinert2018_EPJA54-86}, partly for computational tractability. Chiral Hamiltonians have enabled accurate descriptions of nuclei across the nuclear chart~\cite{Stroberg2021_PRL126-022501,Hu2022_NatPhys18-1196,Soma2020_PRC101-014318,Maris2022_PRC106-064002}. Residual dependence on regulator and resolution scales, however, remains a persistent source of theoretical uncertainty~\cite{Roth2011_PRL107-072501,Hergert2013_PRC87-034307,Hebeler2013_PRC87-031302,Carbone2013_PRC88-044302,Roth2014_PRC90-024325,Binder2014_PLB736-119,Epelbaum2019_PRC99-024313,Carlsson2016_PRX6-011019,Lonardoni2018_PRL120-122502,Drischler2019_PRL122-042501,Tews2025_PRResearch7-033024,Maris2021_PRC103-054001,Lonardoni2020_PRResearch2-022033}. The origin of these differences remains unclear. Moreover, uncertainties in these predictions often motivate tuning interaction parameters to heavier nuclei and nuclear matter~\cite{Ekstrom2015_PRC91-051301,Shirokov2016_PLB761-87,Logoteta2016_PLB758-449,Drischler2019_PRL122-042501,Hoppe2019_PRC100-024318,Huther2020_PLB808-135651,Jiang2024_PRC109-L061302,Elhatisari2024_Nature630-59,Niu2025_PRL135-222504,Hildenbrand2026_PRL136-062501,Elhatisari2026_arXiv2610-20072}, further obscuring essential differences in EFT regularization schemes. Whether few-body calibration alone can yield unique, cutoff-independent predictions for heavier nuclei remains an open question central to the predictive power of nuclear \textit{ab initio} calculations based on EFT.

In this Letter, we show that widely adopted Galilean-invariant regulators for contact interactions can conflict with Wilsonian RG evolution and introduce cutoff sensitivity. Wilsonian momentum-shell analysis reveals that a consistent RG flow requires the same regulator to act on all single-particle momentum modes entering EFT vertices, so that variations from different operators cancel exactly, as in lattice QCD and NRQCD~\cite{Kogut1983_RMP55-775,Lepage1992_PRD46-4052,Hashimoto2004_ARNPS54-451,Kronfeld2012_ARNPS62-265,Aoki2022_EPJC82-869}. This field-operator regularization can break Galilean invariance, but the symmetry can be restored using established lattice-field-theory methodologies such as Symanzik improvement schemes. Within this scheme, we numerically demonstrate cutoff stability for binding energies in finite nuclei and compare with Galilean-invariant schemes in symmetric nuclear matter. Formal details and exactly solvable examples are given in Ref.~\cite{Lu2026_InPreparationII}.

\textit{Wilsonian RG}---For finite cutoff variations, the Wilsonian RG transformation based on momentum-shell integration provides the starting point for establishing cutoff independence~\cite{Wilson1971_PRB4-3174,Wilson1971_PRB4-3184,Wilson1974_PhysRept12-75}. 
Here, we focus on contact interactions, while the pion-exchange potentials can be studied along the same lines.
To illustrate the idea, we consider pionless EFT for nonrelativistic nucleons,
\begin{equation}
\mathcal{L} = \Psi^\dagger \left(i\partial_t + \frac{\nabla^2}{2M}\right) \Psi - \frac{C_2}{2}(\Psi^\dagger \Psi)^2 - \frac{C_3}{6}(\Psi^\dagger \Psi)^3 + \cdots, \label{eq:Lagrangian}
\end{equation}
where $\Psi$ and $\Psi^\dagger$ are conjugate Grassmann fields, $M$ is the nucleon mass, and $C_{2,3}$ denote the low-energy constants (LECs). The ellipsis represents spin-isospin-dependent, derivative and higher-body terms. In the Wilsonian RG scheme, the partition function reads
\begin{equation}
Z_\Lambda = \int \mathcal{D}\Psi_\Lambda\mathcal{D}\Psi^\dagger_\Lambda \exp\left[i\int dtd^3x \mathcal{L}(\Psi_\Lambda, \Psi^\dagger_\Lambda)\right], \label{eq:partition}
\end{equation}
where the subscript $\Lambda$ restricts the path integral to single-particle modes with absolute momenta $|p| \le \Lambda$. We refer to this regularization scheme as the \textit{absolute-momentum} regulator. To transition to a lower cutoff $\Lambda^\prime$, field variables are decomposed into slow and fast modes, $\Psi_\Lambda = \Psi_{\Lambda^\prime} + \delta \Psi$. Integrating out the fast modes within the momentum shell yields an effective Lagrangian of the same form as Eq.~(\ref{eq:Lagrangian}), but with scale-dependent couplings $M^\prime, C_{2,3}^\prime$. Low-momentum observables remain invariant under this transformation, provided all symmetry-allowed operators are retained.
An alternative derivation of the RG flow using unitary transformations based on similarity RG~\cite{Glazek1993_PRD48-5863,Glazek1994_PRD49-4214,Wegner1994_AnnPhys3-77,Bogner2007_PRC75-061001} can be found in Ref.~\cite{Lu2026_InPreparationII}. 
% Althernatively, the RG flow can also be derived using unitary transformations within the similarity RG framework, see Ref.~\cite{Lu2026_InPreparationII}. 

A subtlety arises because the single-particle momentum cutoff in Wilsonian coarse graining selects a preferred reference frame against which the momenta are measured and thus breaks
Galilean invariance. 
Existing Wilsonian RG studies of nuclear EFT have largely focused on
NN scattering in the center-of-mass frame, where boost dependence is
not directly probed. In many-body applications, interactions are
commonly regulated in \textit{relative momenta} to preserve Galilean
invariance. The resulting correlations among single-particle
modes, however, obstruct a direct implementation of momentum-shell integration and obscure
the mapping between successive cutoffs. A universal regulator acting
directly on \textit{absolute single-particle momenta} instead provides
a natural implementation of Wilsonian coarse graining, at the cost of explicit Galilean-invariance breaking. The resulting artifacts can nevertheless be systematically canceled in low-energy observables by including the required symmetry-restoring operators and matching their coefficients order by order~\cite{Li2018_PRC98-044002, Klein2018_EPJA54-233, Li2019_PRC99-064001, Shi2026_PLB874-140303}, in analogy with Symanzik improvement in lattice field theories~\cite{Symanzik1983_NPB226-187,Symanzik1983_NPB226-205,Luscher1985_CMP97-59}.

Galilean-invariant energy-dependent regulators have been considered
in functional RG formulations~\cite{Floerchinger2014_NPA927-119},
but are not directly applicable to nuclear many-body methods based
on energy-independent Hamiltonians.
Here, we focus on a purely spatial momentum-shell construction with
a uniform cutoff on single-particle modes, as in
Eq.~(\ref{eq:partition}), and systematically investigate its many-body
consequences.
Our calculations show that this regulator suppresses high-momentum
single-particle states more strongly than widely used Galilean-invariant
\textit{relative-momentum} regulators, even when the cutoff, LECs, and
center-of-mass NN phase shifts are identical.
This construction provides a consistent Wilsonian formulation of
nuclear EFT, yielding many-body predictions that differ qualitatively
from those of conventional schemes.

%that prioritize Galilean invariance when regulating contact interactions.

% The resulting reduction in nuclear-matter overbinding points to an alternative saturation mechanism. Together, these findings support a consistent Wilsonian formulation of nuclear EFT with qualitatively different many-body behavior from conventional approaches that prioritize Galilean invariance when regulating contact interactions.

\textit{Galilean symmetry breaking}---Regulator-induced symmetry breaking is familiar from lattice QCD, where violations of continuum spacetime symmetries are encoded in irrelevant operators suppressed by powers of the lattice spacing~\cite{Kogut1983_RMP55-775,Davoudi2012_PRD86-054505}. Similarly, our \textit{absolute-momentum} regulator induces Galilean-invariance-breaking operators that are irrelevant in the RG sense, with low-energy contributions suppressed by positive powers of $k/\Lambda$, where $k$ denotes a characteristic nucleon momentum. The leading terms are two-body operators that depend on the pair's total momentum. They vanish in the two-body center-of-mass frame and therefore leave the calibration to NN scattering data unchanged. In nuclei with $A\geq3$, however, an interacting pair generally carries nonzero total momentum, allowing these operators to contribute to binding energies and other observables.

According to Wilsonian RG, the low-energy effects of irrelevant operators can be absorbed into redefinitions of the relevant and marginal couplings, up to residual corrections suppressed by positive powers of $k/\Lambda$. This equivalence becomes asymptotically exact as $\Lambda$ increases at fixed low-energy scales~\cite{Schwartz2014_CUP}. We therefore omit explicit symmetry-restoring counterterms and incorporate their effects through a recalibration of the three-body interactions. The convergence toward results obtained with explicit symmetry restoration is examined for realistic light nuclei in Ref.~\cite{Shi2026_PLB874-140303} and for an exactly solvable one-dimensional bosonic EFT in Ref.~\cite{Lu2026_InPreparationII}.

\begin{figure}[h]
    \centering
        \includegraphics[width=1\linewidth]{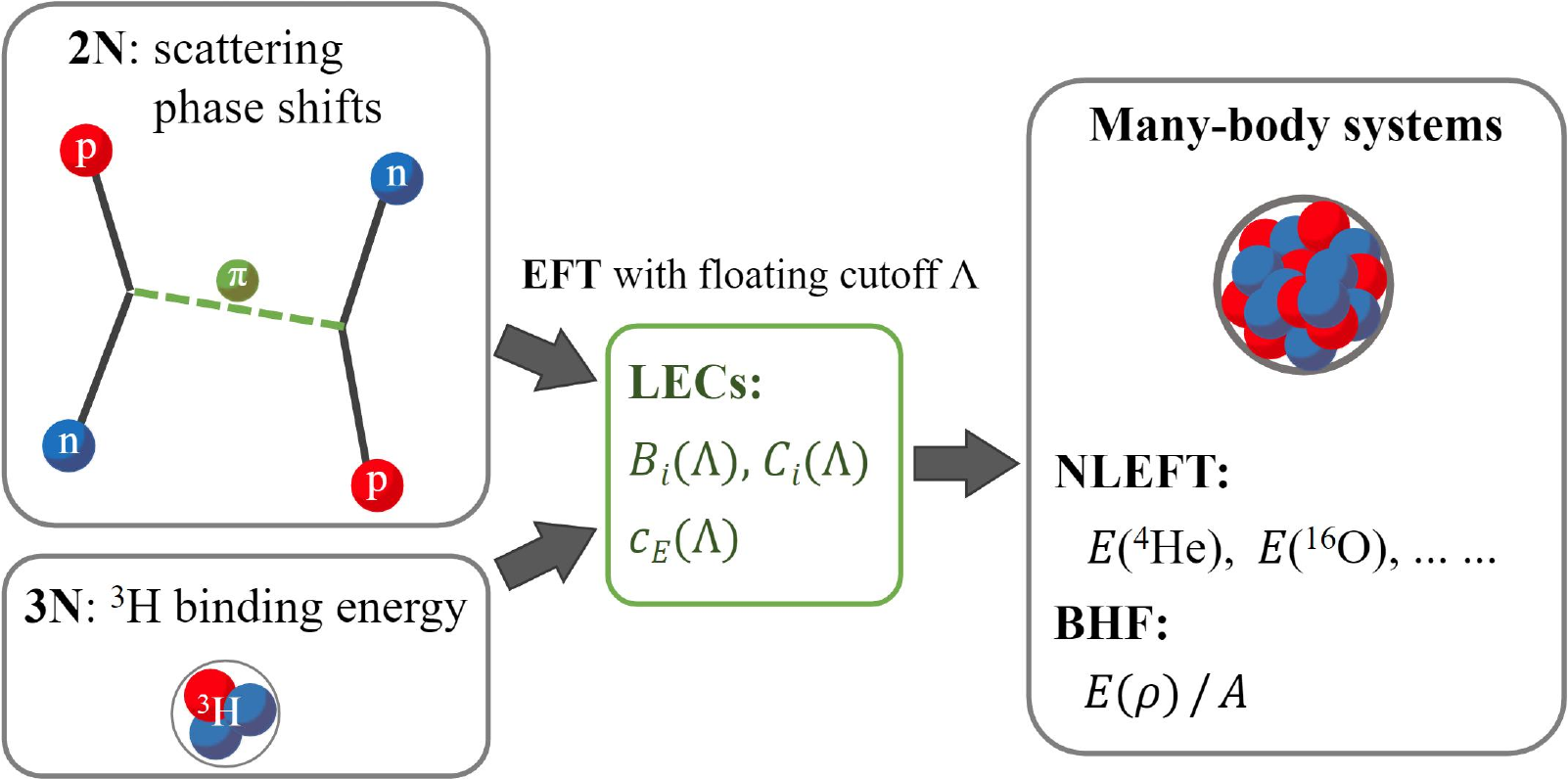}
        \caption{Schematic workflow. Contact LECs are calibrated using neutron-proton scattering data and the triton binding energy, then applied to many-body systems to predict binding energies. The calculation is repeated for different $\Lambda$.}   
        \label{fig:workflow}
\end{figure}

\textit{Method}---The overall workflow of this study is sketched in Fig.~\ref{fig:workflow}. We construct a family of chiral interactions with variable momentum cutoff $\Lambda$ by fitting to two- and three-nucleon data. The resulting interactions are then used as input for many-body calculations. Finite nuclei are computed using the Nuclear Lattice Effective Field Theory (NLEFT)~\cite{Lee2004_PRC70-014007,Lee2005_PRC72-024006,Lee2009_PPNP63-117,Lahde2019_LNP957-1}, while the Brueckner--Hartree--Fock (BHF) method is applied to solve the symmetric nuclear matter (SNM)~\cite{Day1967_RMP39-719,Haftel1970_NPA158-1,Song1998_PRL81-1584}.

We employ the next-to-next-to-leading-order (N$^2$LO) chiral interaction~\cite{Lu2022_PRL128-242501,Liu2025_EPJA61-85,Wu2025_PRC112-014009,Zhang2025_PRD111-036002,Wang2025_PRC112-025502,Shi2026_PLB874-140303}, which consists of both two- and three-nucleon forces (2NFs and 3NFs):
\begin{align}
 V_{{\rm 2N}}= & \left[B_{1}+B_{2}(\bm{\sigma}_{1}\cdot\bm{\sigma}_{2})+C_{1}q^{2}+C_{2}q^{2}(\bm{\tau}_{1}\cdot\bm{\tau}_{2})\right.\nonumber \\
 & +C_{3}q^{2}(\bm{\sigma}_{1}\cdot\bm{\sigma}_{2})+C_{4}q^{2}(\bm{\sigma}_{1}\cdot\bm{\sigma}_{2})(\bm{\tau}_{1}\cdot\bm{\tau}_{2})\nonumber \\
 & +C_{5}\frac{i}{2}(\bm{q}\times\bm{k})\cdot(\bm{\sigma}_{1}+\bm{\sigma}_{2})+C_{6}(\bm{\sigma}_{1}\cdot\bm{q})(\bm{\sigma}_{2}\cdot\bm{q})\nonumber \\
 & \left.+C_{7}(\bm{\sigma}_{1}\cdot\bm{q})(\bm{\sigma}_{2}\cdot\bm{q})(\bm{\tau}_{1}\cdot\bm{\tau}_{2})\right]f_{{\rm abs}}(p_1^\prime, p_1, p_2^\prime, p_2)\nonumber \\
 & -\frac{g_{A}^{2}f_{\pi}(q^{2})}{4F_{\pi}^{2}}\left[\frac{(\bm{\sigma}_{1}\cdot\bm{q})(\bm{\sigma}_{2}\cdot\bm{q})}{q^{2}+M_{\pi}^{2}}+C_{\pi}^{\prime}\bm{\sigma}_{1}\cdot\bm{\sigma}_{2}\right](\bm{\tau}_{1}\cdot\bm{\tau}_{2}),  \nonumber \\
 V_{3\mathrm{N}} = & -\frac{3c_E}{F_\pi^4 \Lambda_\chi} 
 f_\mathrm{abs}(p_1^\prime, p_1, p_2^\prime, p_2, p_3^\prime, p_3),
\label{eq:V2N}
\end{align}
where $\bm{p}_{1,2,3}$ and  $\bm{p}^\prime_{1,2,3}$ are 
incoming and outgoing momenta of individual nucleons,
with $\bm{p}' = (\bm{p}'_1 - \bm{p}'_2)/2$
$\bm{p} = (\bm{p}_1 - \bm{p}_2)/2$ being the relative momenta, 
$\bm{q} = \bm{p}^\prime-\bm{p}$ and 
$\bm{k} = (\bm{p}^\prime+\bm{p})/2$ 
are momentum transfers,
    %with $\bm{p}^\prime = (\bm{p}_1^\prime - \bm{p}_2^\prime)/2$ %and $\bm{p} = (\bm{p}_1 - \bm{p}_2)/2$ 
    % the relative momenta, 
$\bm{\sigma}_{1,2}$ and $\bm{\tau}_{1,2}$ are Pauli matrices 
for spin and isospin, respectively.  
For the one-pion exchange potential (OPEP), $g_A$, $F_\pi$ and $M_\pi$  are the axial-vector coupling constant, pion decay constant and pion mass, respectively, with $C_\pi^\prime$  a constant counterterm and $f_\pi(q^2)$ a local regulator for pion momentum~\cite{Reinert2018_EPJA54-86}. In the 3NF, $\Lambda_\chi$ is the chiral symmetry breaking scale~\cite{Epelbaum2010_EPJA45-335} and $c_E$ is a dimensionless 
adjustable parameter.  
All contact terms are regulated by suppressing high \emph{absolute} single-particle momenta,
\begin{equation}\label{eq:fSP}
	f_\mathrm{abs}(p_1^\prime,p_1,\dots,p_A^\prime,p_A) = \prod_{i=1}^A
    \exp\left(-\frac{p^{\prime 2n}_i + p^{2n}_i}{2\Lambda^{2n}}
    \right), 
\end{equation}
with $\Lambda$ the momentum cutoff.
Here we take $n = 3$~\cite{Lu2022_PRL128-242501,Liu2025_EPJA61-85,Wu2025_PRC112-014009,Zhang2025_PRD111-036002,Wang2025_PRC112-025502,Shi2026_PLB874-140303}.
Such a regulator can be viewed as a continuum counterpart of the lattice regulator, which imposes a uniform cutoff on all field momentum modes.
For finite nuclei, a perturbative static Coulomb force is included. 
Further details of the methods, interactions and dependencies of the results on OPEP regularization are provided in the \hyperref[sec:supplemental-material]{Supplemental Material}.

In the main text, we keep the OPEP fixed and concentrate on the renormalization of the short-range contact terms. 
The LECs $B_{1-2}$ and $C_{1-7}$ are determined by fitting to empirical NN scattering phase shifts~\cite{Stoks1993_PRC48-792} up to a relative momentum of $p_{\mathrm{rel}} \leq 200~\mathrm{MeV}$~\cite{Lu2016_PLB760-309}, while the three-body LEC $c_E$ is adjusted to the experimental triton binding energy $E(^3\mathrm{H}) = -8.482~\mathrm{MeV}$.
In this momentum range, where 
$M_\pi \simeq p_{\mathrm{rel}}<2M_\pi$, the two-pion exchange potentials are analytic and can be effectively absorbed into the contact terms~\cite{Li2018_PRC98-044002,Wu2025_PRC112-014009,Liu2025_EPJA61-85}.

Here, we compare the \emph{absolute}-momentum regulator $f_{\rm abs}$ defined in Eq.~(\ref{eq:fSP}) with the widely used Galilean-invariant nonlocal regulator $f_{\rm rel}$~\cite{Entem2003_PRC68-041001,Epelbaum2005_NPA747-362,Epelbaum2015_EPJA51-53,Entem2017_PRC96-024004,Reinert2018_EPJA54-86}.
For the 2NFs, these regulators read
\begin{align}
f_\mathrm{abs}
&=\exp\!\left[-\frac{p_1^{\prime 2n}+p_1^{2n}+p_2^{\prime 2n}+p_2^{2n}}{2\Lambda^{2n}}\right], \label{fSP2}\\
f_\mathrm{rel}
&=\exp\!\left[-\frac{|\bm{p}_1^\prime-\bm{p}_2^\prime|^{2n}+|\bm{p}_1-\bm{p}_2|^{2n}}{(2\Lambda)^{2n}}\right]. \nonumber
\end{align}
The regulators coincide in the two-nucleon (2N) center-of-mass frame, where $\bm{p}_1=-\bm{p}_2$ and $\bm{p}_1^\prime=-\bm{p}_2^\prime$. Fits to the same 2N phase shifts therefore yield identical LECs at a given $\Lambda$.
In many-body systems, however, $f_{\mathrm{abs}}$ suppresses configurations with large individual momenta, whereas $f_{\mathrm{rel}}$ constrains only the relative momenta, leaving the total momentum unrestricted.
For the same $\Lambda$ and LECs, $f_{\mathrm{abs}}$ thus suppresses intermediate states more strongly, leading in our calculations to reduced in-medium scattering amplitudes and significantly weaker many-body binding relative to $f_{\mathrm{rel}}$.

    \begin{figure}[h] 
        \centering  
        \includegraphics[width=1\linewidth]{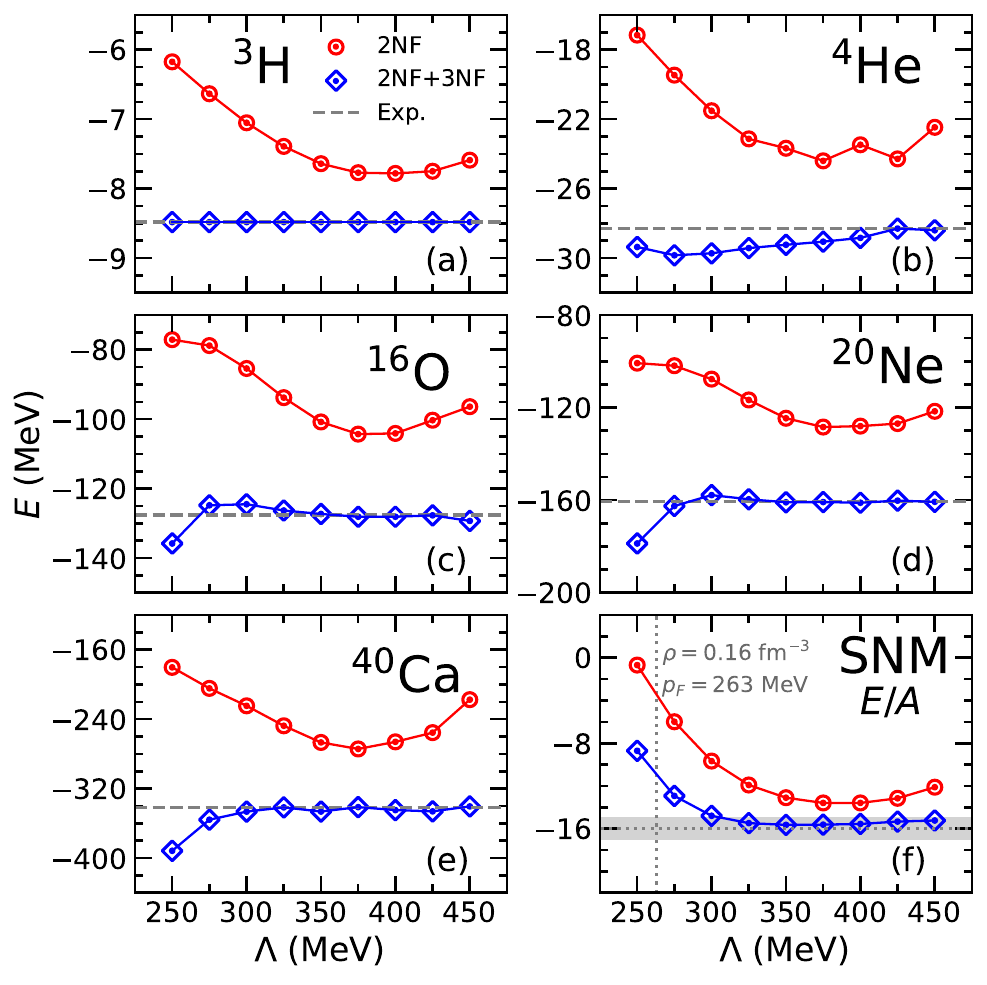} 
         \caption{(a-e) Binding energies of selected nuclei as functions of the cutoff $\Lambda$.
         Circles and diamonds denote results without and with 3NF, respectively.
         Statistical errors are smaller than the symbols and not shown.
         Dashed lines mark the experimental values.
         (f) Binding energy per nucleon for symmetric nuclear matter at empirical saturation density $\rho_0 = 0.16$~fm$^{-3}$.
         Vertical line represents the corresponding Fermi momentum. The empirical binding 
         energy per nucleon $E/A = -16\pm 1$~MeV
         is indicated by the horizontal band.
         Numerical results are provided in the \hyperref[sec:supplemental-material]{Supplemental Material}.
                  }\label{fig:LamRun}
    \end{figure}

\textit{Finite nuclei}---To investigate regulator dependence, the momentum cutoff $\Lambda$ is tuned from $250$ to $450$ MeV.
Fig.~\ref{fig:LamRun}(a--e) displays the binding energies of nuclei ranging from $^{3}$H to $^{40}$Ca as functions of the momentum cutoff $\Lambda$.
To elucidate the role of 3NFs, we compare the full results with calculations employing 2NFs alone.
The 2NF-only results systematically underbind these nuclei and exhibit a pronounced dependence on $\Lambda$, varying by about $20\%$--$30\%$ in the considered $\Lambda$-interval.
On the other hand, for the full 2NF+3NF interaction, the triton binding energy is reproduced by construction through our fitting procedure. In comparison, the predictions for heavier nuclei [Fig.~\ref{fig:LamRun}(b--e)] demonstrate clear RG invariance.
Particularly, for $\Lambda \geq 300~\mathrm{MeV}$, the binding energies remain essentially stable, with variations of only a few MeV, while simultaneously aligning with experimental data. Specifically, the $^{40}$Ca binding energy is predicted to be $-343.9(27)(25)$~MeV, where the first uncertainty accounts for the $\Lambda$ variation between $300$ and $450$~MeV and the second represents the many-body statistical error from NLEFT. 
This result is in agreement with the experimental value of $-342.0$~MeV. 
Notably, the cutoff dependence of all nuclei observed in the 2NF-only sector is consistently absorbed by incorporating a single 3NF counterterm. 
We further note that the residual overbinding in $^4\mathrm{He}$ 
is consistent with our previous studies and can be eliminated by explicitly including the Galilean invariance restoration terms~\cite{Shi2026_PLB874-140303}.

\begin{figure}[h]
       \centering  
       \includegraphics[width=\columnwidth] {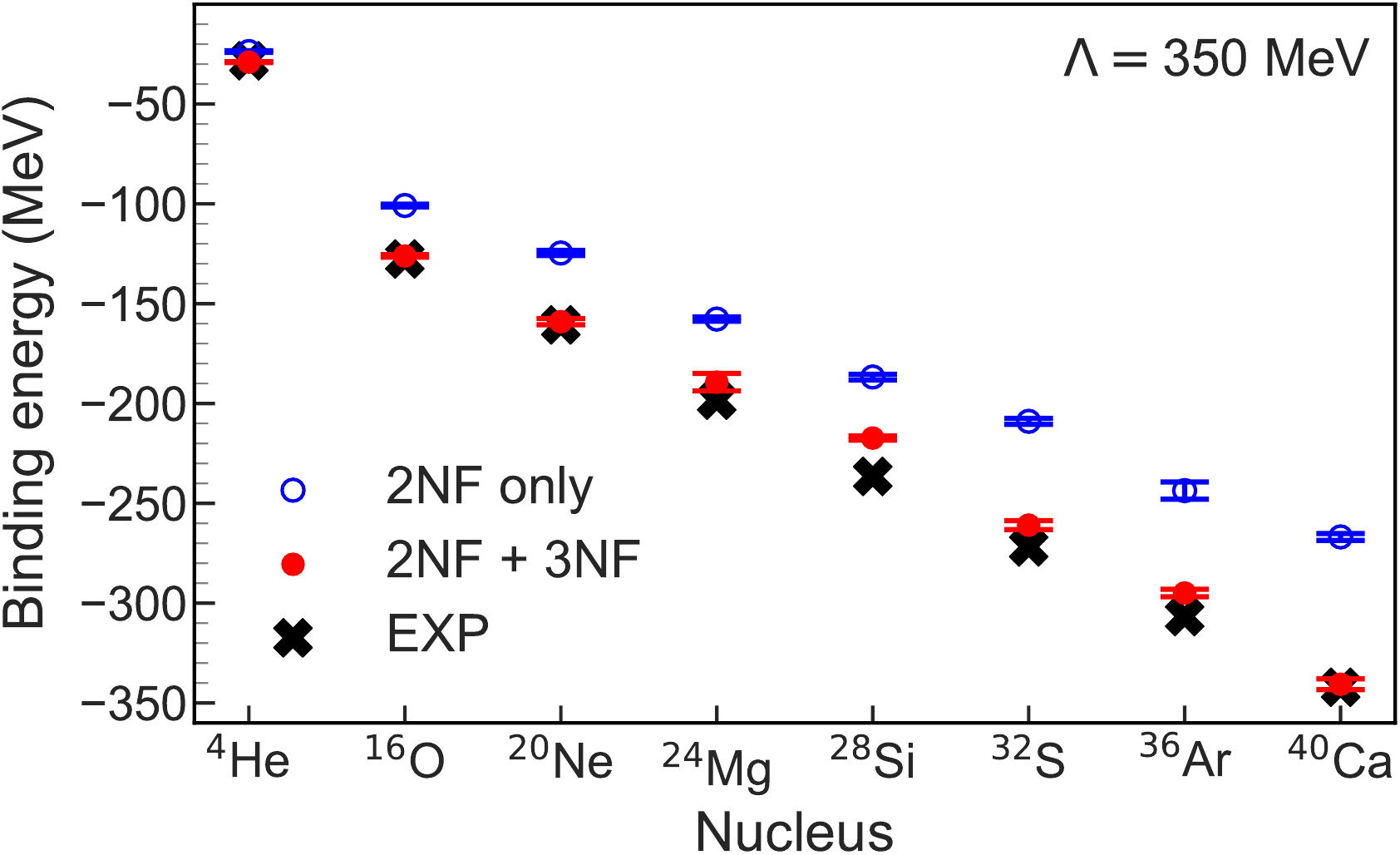}
       \caption{Binding energies calculated with two-body (2NF) and two- plus three-body forces (2NF + 3NF). }
        \label{fig:nucls350}
    \end{figure}

Fig.~\ref{fig:nucls350} displays the binding energies for several even-even nuclei computed at a fixed cutoff 
$\Lambda = 400~\mathrm{MeV}$.
The 2NF-only results systematically underbind these nuclei by 
approximately $20$\%, whereas including 3NFs significantly improves the agreement with 
experimental data.

\textit{Nuclear matter}---SNM saturation provides a stringent test of nuclear interactions. BHF calculations with phase-shift-equivalent 2NFs typically yield saturation points along the Coester line, away from the empirical point~\cite{Coester1970_PRC1-769,Furnstahl2001_NPA689-846,Li2006_PRC74-047304}. Soft chiral 2NFs exhibit similar deficiencies in BHF and many-body perturbation theory, often producing substantial overbinding and unrealistic saturation~\cite{Hebeler2011_PRC83-031301,Li2012_PRC85-064002,Hu2017_PRC96-034307,Drischler2019_PRL122-042501}. Recovering empirical saturation then often requires fine-tuning 3NF LECs, such as $c_D$ and $c_E$, to balance 2NF attraction.

\begin{figure}[h]
        \centering
        \includegraphics[width=1\linewidth]{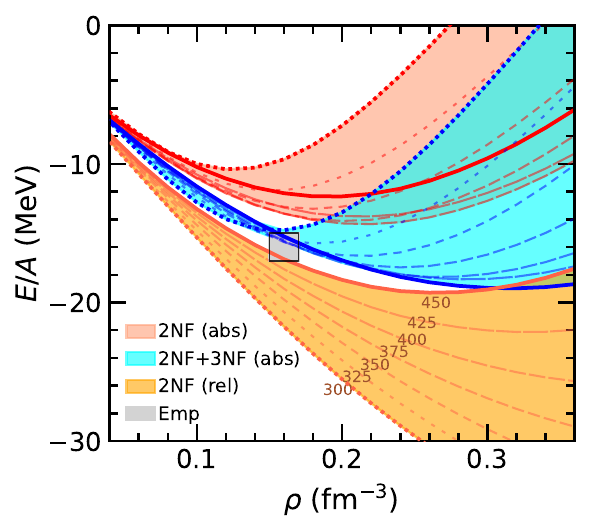}
        \caption{Binding energy per nucleon of symmetric 
        nuclear matter.
        Red (Top) and blue (middle) lines denote 
        results of \emph{absolute}-momentum regulator (abs) 
        without and with 3NF, respectively.
        Orange (Bottom) lines indicate results calculated 
        using 2NF with the same LECs and conventional 
        \emph{relative}-momentum regulator (rel).
        The uncertainty bands are generated by varying 
        cutoff $\Lambda$ from $300$~MeV (dotted lines) 
        to $450$~MeV (solid lines).
        Grey patch shows the empirical saturation point.
        }\label{fig:snm_eos}
    \end{figure}

Fig.~\ref{fig:LamRun}(f) shows the cutoff dependence of $E/A$ at $\rho_0=0.16\,\mathrm{fm}^{-3}$. For $\Lambda\geq300\,\mathrm{MeV}$, the full $2N+3N$ interaction yields a nearly cutoff-independent $E/A$ consistent with the empirical saturation energy $E/A\approx-16\,\mathrm{MeV}$. Deviations emerge at $\Lambda\leq275\,\mathrm{MeV}$, where the cutoff approaches or falls below the characteristic Fermi momentum $p_F\approx263\,\mathrm{MeV}$, signaling a loss of scale separation and the breakdown of the EFT. Fig.~\ref{fig:snm_eos} compares the equations of state (EOS) with and without 3NFs for $\Lambda=300$--$450\,\mathrm{MeV}$, with the bands indicating cutoff variation. The 2NF-only band broadens rapidly with density, whereas including 3NFs yields a narrow, nearly cutoff-independent band below $\rho_0$. Above $\rho_0$, the band broadens as characteristic momenta increasingly probe high-momentum scales beyond the range for which our framework is calibrated ($p_{\rm rel}\leq200\,\mathrm{MeV}$). Reproducing the empirical saturation point and improving the high-density EOS beyond $\rho_0$ will require both higher-precision interactions and many-body treatments beyond BHF. Our results for nuclei and SNM both probe the low-cutoff limit and support the expectation that an EFT provides robust predictions only when the characteristic momentum, typically estimated by the average or Fermi momentum, lies well below the cutoff.

Fig.~\ref{fig:snm_eos} directly compares the 2NF-only EOSs computed with $f_{\rm abs}$ and $f_{\rm rel}$ using identical LECs and the same BHF implementation. Throughout the investigated cutoff range, $f_{\rm rel}$ produces substantial overbinding and unrealistic saturation, consistent with previous BHF calculations using relative-momentum regulators~\cite{Li2006_PRC74-047304,Li2012_PRC85-064002,Hu2017_PRC96-034307}. In contrast, $f_{\rm abs}$ yields underbound EOSs with well-defined saturation minima. Since overbinding in SNM often accompanies that in heavier nuclei, these results suggest an alternative to balancing excessive 2NF attraction with large repulsive 3NF contributions. The remaining underbinding leaves room for attractive 3NFs, consistent with the 2NF-only triton result in Fig.~\ref{fig:LamRun}(a).

\textit{Summary and perspective}---Inspired by the Wilsonian RG principles and renormalization of lattice field theories, we introduce a single-particle regularization scheme for nuclear EFT contact interactions, yielding nearly cutoff-independent binding energies for nuclei up to $^{40}$Ca with all couplings calibrated only to $A\leq3$ data. The predicted energies agree well with experiment, while the nuclear-matter EOS differs substantially from that obtained with conventional relative-momentum regulators. These results provide a practical route to improved RG invariance in nuclear \textit{ab initio} calculations. Further tests with more general chiral interactions and other nonrelativistic EFTs will clarify the scope of this framework, with potential applications in hadronic physics, quantum chemistry, and ultracold atomic gases.

\textit{Acknowledgements.}---We thank members of the Nuclear Lattice 
Effective Field Theory Collaboration for insightful discussions.
B. L. thanks Dean Lee for critical reading of the manuscript.
C. W. thanks Prof. X. L. Shang for enlightening discussions.
This work was supported by the Science Challenge Project (No. TZ2025012), NSAF (No. U2330401), and the National Natural Science Foundation of China under Grant Nos. 12275259 and 12547105.

\textit{Data availability.} The data are available from the authors upon 
request.

%\newpage

\clearpage

% Independent Supplemental Material numbering with unique hyperlink anchors.
\setcounter{equation}{0}
\setcounter{figure}{0}
\setcounter{table}{0}
\renewcommand{\theequation}{S\arabic{equation}}
\renewcommand{\thefigure}{S\arabic{figure}}
\renewcommand{\thetable}{S\arabic{table}}
\renewcommand{\theHequation}{supp.\arabic{equation}}
\renewcommand{\theHfigure}{supp.\arabic{figure}}
\renewcommand{\theHtable}{supp.\arabic{table}}
\setcounter{secnumdepth}{-2}
\setcounter{tocdepth}{-2}
\allowdisplaybreaks[4]
\renewcommand{\no}[1]{:\!#1\!:}

\onecolumngrid
\phantomsection
\label{sec:supplemental-material}
\pdfbookmark[0]{Supplemental Material}{supplemental-material}
\begin{center}
  {\large\bfseries Supplemental Material}
\end{center}
\twocolumngrid

This appendix presents the computational methods and numerical details, including the determination of the low-energy constants and the numerical setups for the NLEFT and BHF calculations. Previously published details, cited in the main text, are included here for completeness.

We provide the numerical data underlying the figures in the main text, together with results for nuclei up to $^{16}$O and for nuclear matter with the Galilean-invariance-restoration corrections included. Finally, we discuss possible extensions to higher-order chiral interactions.

\section*{Determination of low-energy constants}
    We use the average nucleon mass $M_{\rm N}=938.92$~MeV for both protons and neutrons.
    The two-nucleon force (2NF) adopted in this work consists of 
    the long-range one-pion-exchange potential
    (OPEP, $V_{1\pi}$) and short-range contact interactions $V_{\rm ct}$.
    We define $\bm p=(\bm p_1-\bm p_2)/2$,
    $\bm p'=(\bm p_1'-\bm p_2')/2$,
    $\bm q=\bm p'-\bm p$, and $\bm k=(\bm p'+\bm p)/2$.
    This momentum convention follows Ref.~\cite{Liu2025_EPJA61-85}.
    The common single-particle regulator multiplies the contact terms,
    whereas the OPEP uses only the local pion-momentum regulator specified below.
    The OPEP reads 
    \begin{equation}\label{OBEP}
        V_{1\pi} = -\bm{\tau}_{1}\cdot\bm{\tau}_{2}
        \frac{g_A^2 f_{\pi}(q^2)}{4F_\pi^2}\left[
        \frac{(\bm{\sigma}_1\cdot\bm{q})(\bm{\sigma}_2\cdot\bm{q})}
        {q^2+M_\pi^2}+C_\pi'\bm{\sigma}_1\cdot\bm{\sigma}_2\right].
    \end{equation}
    Here the axial-vector coupling constant, decay constant and 
    pion mass are $g_A=1.287$, $F_\pi = 92.2$~MeV and 
    $M_\pi = 134.980$~MeV. 
    The OPEP is regulated using the semilocal scheme~\cite{Reinert2018_EPJA54-86},
    in which
    $f_\pi(q^{2})=\exp[-(q^{2}+M_{\pi}^{2})/\Lambda_\pi^{2}]$, 
    $\Lambda_\pi=300$~MeV and the counterterm coefficient is
    \begin{align*}
    C'_\pi={}&-\frac13\Biggl[1-\frac{2M_\pi^2}{\Lambda_\pi^2}\\
    &\quad+2\sqrt\pi\frac{M_\pi^3}{\Lambda_\pi^3}
    e^{M_\pi^2/\Lambda_\pi^2}
    \operatorname{erfc}\left(\frac{M_\pi}{\Lambda_\pi}\right)\Biggr].
    \end{align*}    
    
    \begin{figure*}
    \centering
        \includegraphics[width=1\linewidth]{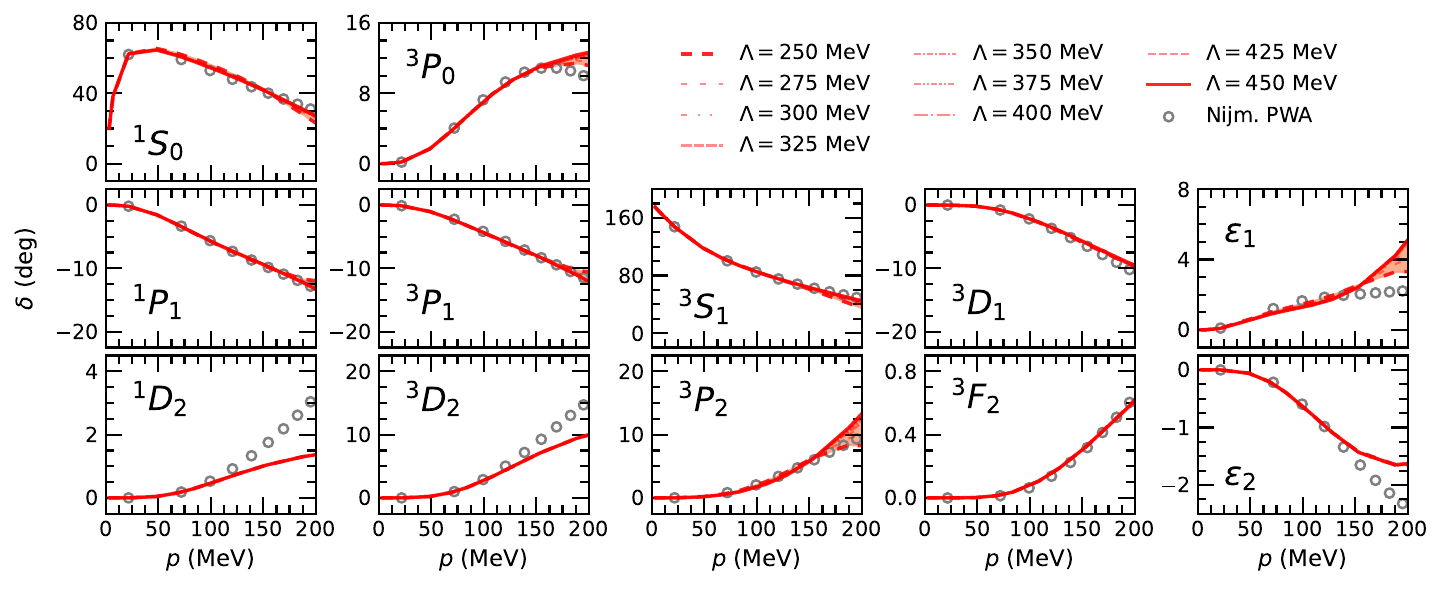}
        \caption{Nucleon-nucleon scattering phase shifts calculated 
        using the lattice chiral force with $\Lambda=250,~275,~\cdots,~425,~450$ MeV. 
        The Nijmegen partial-wave phase shifts (empty circles) are 
        also plotted for comparison. }
        \label{fig:fishes}
    \end{figure*} 
    
    The contact term $V_{\rm ct}$ contains
    \begin{subequations}\label{Vct:LECs}
    \begin{align}
        V_{Q^0} &=  B_1+B_2\bm{\sigma}_{1}\cdot\bm{\sigma}_{2}
            \\  \nonumber
        V_{Q^2} &= q^2\left[ C_1+C_2 \bm{\tau}_1\cdot\bm{\tau}_2+
        C_3 \bm{\sigma}_1\cdot\bm{\sigma}_2 \right.
            \\ \nonumber
        &+ \left. C_4 (\bm{\sigma}_1\cdot\bm{\sigma}_2)
        (\bm{\tau}_1\cdot\bm{\tau}_2)\right]
        +C_5 i(\bm{q}\times\bm{k})\cdot(\bm{\sigma}_1+\bm{\sigma}_2)/2
            \\
        &  +(\bm{\sigma}_{1}\cdot\bm{q})
        (\bm{\sigma}_{2}\cdot\bm{q})
        (C_6 + C_7 \bm{\tau}_1\cdot\bm{\tau}_2).    
    \end{align}
    \end{subequations} 
    To obtain the low-energy constants (LECs) including
    $B_1,~B_2$ and $C_{1\text{--}7}$,
    we first determine the spectroscopic LECs in each partial wave.
    The corresponding linear relations are
    \begin{subequations}\label{LECs:PWD}
    \begin{align}
    B_{{}^1S_0}&=B_1-3B_2,\\
    B_{{}^3S_1}&=B_1+B_2,\\
    C_{{}^1S_0}&=C_1+C_2-3C_3-3C_4-C_6-C_7,\\
    C_{{}^3P_0}&=-\frac23(C_1+C_2+C_3+C_4-C_5)\nonumber\\
      &\quad+2(C_6+C_7),\\
    C_{{}^1P_1}&=-\frac23(C_1-3C_2-3C_3+9C_4-C_6)\nonumber\\
      &\quad-2C_7,\\
    C_{{}^3P_1}&=-\frac23(C_1+C_2+C_3+C_4)+\frac{C_5}{3}\nonumber\\
      &\quad-\frac43(C_6+C_7),\\
    C_{{}^3S_1}&=C_1-3C_2+C_3-3C_4+\frac{C_6}{3}-C_7,\\
    C_{SD}&=-\frac{2\sqrt2}{3}C_6+2\sqrt2 C_7,\\
    C_{{}^3P_2}&=-\frac23(C_1+C_2+C_3+C_4)-\frac{C_5}{3}.
    \end{align}
    \end{subequations}
    In the two-body center-of-mass frame, the regulator depends only
    on $p$ and $p'$, and the partial-wave contact terms are
    given by
    \begin{subequations}\label{Vct:Q2}
    \begin{align}
        &\langle S|V_{Q^0}|S\rangle=B_S 
        f_{\rm abs}(p'_1,p_1,p_2',p_2),
            \\
    	& \langle S|V_{Q^2}|S\rangle=C_S~(p'^2 + p^2) 
        f_{\rm abs}(p'_1,p_1,p_2',p_2), 
             \\
        &\langle S|V_{Q^2}|D\rangle= C_{SD}~p^2 
        f_{\rm abs}(p'_1,p_1,p_2',p_2), 
            \\
    	&\langle P|V_{Q^2}|P\rangle = C_P~p'p 
        f_{\rm abs}(p'_1,p_1,p_2',p_2).
    \end{align} 
    \end{subequations} 
    
    These expressions use the normalization of the spectroscopic
    LECs adopted here, with the common angular factor absorbed into
    the partial-wave convention. The sign of $C_{SD}$ fixes the relative
    phase of the $S$- and $D$-wave basis states, and the Hermitian
    conjugate of the $S$--$D$ matrix element contains $p'^2$.

    We compute the scattering phase shifts $\delta$ at a given
    relative momentum $p$ using the lattice spherical wall method 
    with angular momentum projection and auxiliary 
    potential~\cite{Lu2016_PLB760-309}. 
    The spectroscopic LECs are determined by optimizing  
    \begin{equation}
        \chi^2 = \frac{1}{N_p}\sum_p^{N_p}\left[
        \frac{\delta(p)-\delta_\mathrm{Nijm}(p)}
        {\Delta \delta_\mathrm{Nijm}(p)}\right]^2,
    \end{equation} 
    with $\delta_\mathrm{Nijm}$ and $\Delta\delta_\mathrm{Nijm}$ 
    being the Nijmegen partial-wave phase shifts and corresponding
    uncertainties~\cite{Stoks1993_PRC48-792} up to a relative momentum 
    $p=200$~MeV (corresponding to laboratory energy 
    $E_{\rm lab} = 85$ MeV). 
    The $B_i$ and $C_i$ are then obtained by inverting the linear relations in Eqs.~\eqref{LECs:PWD}. 
 
    By construction, the fitted LECs reproduce  
    the $S$- and $P$-wave phase shifts within the calibration 
    range $p \leq 200$ MeV for cutoffs $\Lambda \in [250, 450]$ MeV.
    No independent diagonal $D$- or $F$-wave contact terms occur
    at this order. Uncoupled channels without contact interactions
    are therefore determined by the fixed OPEP, while coupled channels
    can also depend on the contact interactions in their partner channels.
    Our sensitivity analysis favors the adopted value $g_A=1.287$
    throughout the cutoff range considered here.

\begin{figure*}
    \centering
        \includegraphics[width=1\linewidth]{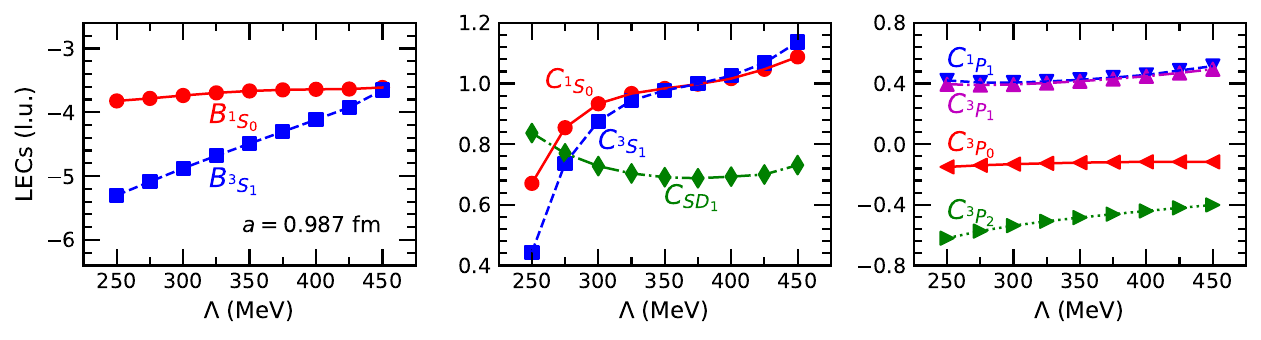}
        \caption{LECs of $V_{\rm ct}$ as functions of the cutoff.
        The reference units are $(200~\mathrm{MeV})^{-2}$ for $B$
        and $(200~\mathrm{MeV})^{-4}$ for $C$.
        The inverse reference scale corresponds to a length
        $a_0=(200~\mathrm{MeV})^{-1}\simeq0.987$~fm.
        The plotted label $C_{SD1}$ denotes $C_{SD}$.}
        \label{fig:lecrunning}
    
\end{figure*}

    The fitted LECs are depicted in 
    Fig.~\ref{fig:lecrunning}.  
    As the cutoff $\Lambda$ is varied from $250$ to $450$~MeV, each 
    spectroscopic LEC evolves continuously to compensate for the 
    truncated high-momentum physics. This smooth running is consistent with the expected
    cutoff dependence of the LECs. The stability of observables after
    refitting provides the corresponding test of renormalization group
    (RG) invariance. 
    
    The interaction omits two-pion-exchange contributions in both
    the two- and three-nucleon sectors. This is an additional approximation.
    Over the low-momentum range considered here, their short-range
    contributions can be partly absorbed into the refitted contact
    LECs, as in Refs.~\cite{Lu2022_PRL128-242501,Liu2025_EPJA61-85}.
    The strong Tjon correlation between the triton and alpha-particle
    binding energies limits their ability to constrain the short-range
    three-nucleon couplings $c_D$ and $c_E$ independently~\cite{tjon1975plb}.
    We therefore retain only the purely contact three-nucleon term
    proportional to $c_E$. At each cutoff, $c_E$ is fixed by reproducing
    the experimental triton energy $E_{\rm Exp.}({}^3\mathrm H)=-8.482$~MeV.

    On the discrete lattice, the triton ground state is obtained 
    via sparse matrix diagonalization using the Lanczos algorithm. 
    To estimate the infinite-volume limit, we use an extrapolation
    ansatz motivated by the three-body result in the unitary limit~\cite{meissner2015prl}
    \begin{equation}\label{3Nextrap}
        E_{^3{\rm H}}(L) = E_{^3{\rm H}}(+\infty) + 
        \frac{\mathcal{C}}{(q_{\rm B} L)^\frac{3}{2}} \exp
        \left(-2q_{\rm B} L/\sqrt{3}\right),
    \end{equation}
    where $E_{^3{\rm H}}(+\infty)$ is the infinite-volume energy and
    $q_{\rm B}=\sqrt{-E_{^3{\rm H}}(+\infty)M_{\rm N}}$.
    The product $q_{\rm B}L$ is understood in consistent lattice units,
    with physical box length $aL$. This functional form is used as a
    fitting ansatz rather than an exact asymptotic relation for a
    physical triton with a bound deuteron subsystem.

    \begin{figure*}
    \centering
        \includegraphics[width=.75\linewidth]{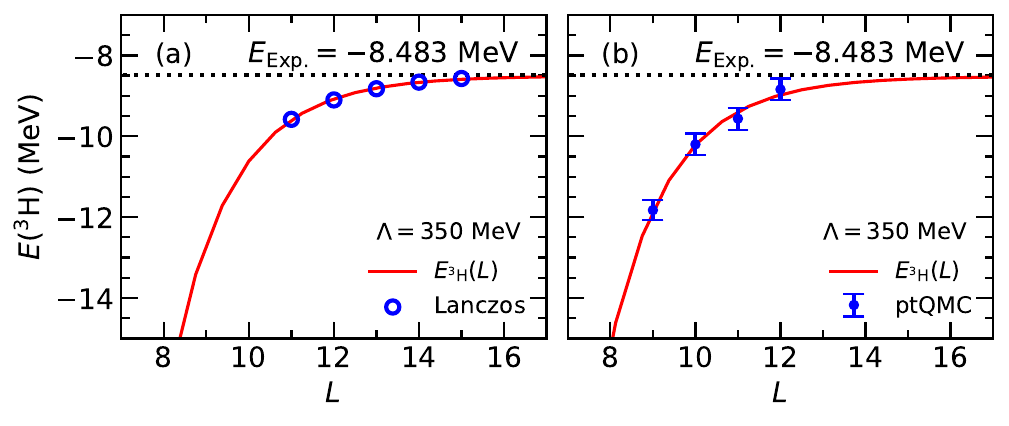}
        \caption{(a) Triton binding energy calculated by
        the Lanczos algorithm with infinite-volume extrapolation.
        (b) Triton binding energy calculated by
        perturbative quantum Monte Carlo method with 
        infinite-volume extrapolation.}
        \label{fig:finvol}
    
\end{figure*} 
    Figure~\ref{fig:finvol}(a) shows the triton energies obtained
    with the Lanczos algorithm in boxes of varying $L$, together with
    the fit to Eq.~\eqref{3Nextrap}. The three-body coupling is calibrated
    to the extrapolated energy. Figure~\ref{fig:finvol}(b) shows the
    corresponding ptQMC calculation~\cite{Lu2022_PRL128-242501}, which agrees with
    the calibrated triton energy within its numerical uncertainty.

\section*{Perturbative quantum Monte Carlo method}

We follow the ptQMC formulation of Ref.~\cite{Lu2022_PRL128-242501} and its
FFT implementation in Ref.~\cite{Liu2025_EPJA61-85}. A Wigner-SU(4)
Hamiltonian is solved nonperturbatively, and the difference between
this reference Hamiltonian and the target interaction is included
through second order. Fully nonperturbative results, where available,
provide benchmarks for this expansion.

\subsection*{Reference Hamiltonian and regulated fields}

We use $\hbar=c=1$. Coordinate-space operator expressions are written in lattice units with $a=1$, and dimensional momenta and couplings are converted to these units before implementation. If not specified, for calculations with $\Lambda\leq400$~MeV, we use $a=(200~\mathrm{MeV})^{-1}\approx0.987$~fm, corresponding to a lattice momentum cutoff $\Lambda_a=\pi/a\approx628$~MeV. For $\Lambda>400$~MeV, we use the finer spacing $a=(250~\mathrm{MeV})^{-1}\approx0.789$~fm, with $\Lambda_a\approx785$~MeV. Thus, $\Lambda<\Lambda_a$ for all cutoffs considered.
We have examined convergence toward the continuum limit by varying $a$ at fixed $\Lambda$. Within the numerical accuracy considered, the predictions typically converge to their continuum-limit values once $a\lesssim2\hbar c/\Lambda$. For consistency, all LECs are reported in units defined by the common reference length $a_0=(200~\mathrm{MeV})^{-1}$, irrespective of the lattice spacing used in the calculation.
In Monte Carlo calculations, we increase the number of lattice sites as the spacing decreases. At $a=(200~\mathrm{MeV})^{-1}$, we use $L=12$ for nuclei with $A<30$ and $L=13$ for $A\geq30$. At $a=(250~\mathrm{MeV})^{-1}$, we use $L=14$ throughout. For Lanczos diagonalization of the triton, we use $L=15$ and account for finite-volume corrections when fitting to the experimental binding energy. Calculations with different box sizes confirm that the results are stable against residual finite-volume effects.

On a periodic
$L^3$ lattice, the reference Hamiltonian is
\begin{equation}\label{eq:SU(4)Hamiltonian}
H_0=K+\frac{C_{\rm SU4}}{2}\sum_{\bm n}\no{\widetilde\rho^2(\bm n)},
\end{equation}
where the colons denote normal ordering and
\begin{equation}\label{eq:kinetic}
K=\sum_{\bm n,\alpha}a^\dagger(\bm n,\alpha)
\frac{-\nabla^2}{2M_{\rm N}}a(\bm n,\alpha).
\end{equation}
The index $\alpha$ labels the four spin-isospin states.
The SU(4) symmetry~\cite{wigner1937pr} makes this interaction
independent of the spin and isospin orientations.
Following Refs.~\cite{lu2019plb,Liu2025_EPJA61-85}, the locally smeared
density is
\begin{align}\label{eq:local-smearing}
\widetilde\rho(\bm n)&=\sum_\alpha
\widetilde a^\dagger(\bm n,\alpha)\widetilde a(\bm n,\alpha)
\nonumber\\
&\quad+s_{\rm L}\sum_{|\bm n'-\bm n|=1}\sum_\alpha
\widetilde a^\dagger(\bm n',\alpha)\widetilde a(\bm n',\alpha).
\end{align}
The nonlocal smearing is implemented by FFT,
\begin{align}\label{NLsmearing}
\widetilde a(\bm n,\alpha)
&=\frac{1}{L^{3/2}}\sum_{\bm p}e^{i\bm p\cdot\bm n}
e^{-p^2/(2\Lambda_{\rm NL}^2)}a(\bm p,\alpha)
\nonumber\\
&=\sum_{\bm n'}g(\bm n-\bm n')a(\bm n',\alpha),
\end{align}
where the momentum sum runs over the lattice Brillouin zone and
\begin{equation}
g(\bm n)=\frac{1}{L^3}\sum_{\bm p}
e^{i\bm p\cdot\bm n}e^{-p^2/(2\Lambda_{\rm NL}^2)}.
\end{equation}

For $a=(200~\mathrm{MeV})^{-1}$, the reference-interaction parameters are $s_{\rm L}=0.182$, $\Lambda_{\rm NL}=300$~MeV, and $C_{\rm SU4}=-3.7912\times10^{-5}$~MeV$^{-2}$. For $a=(250~\mathrm{MeV})^{-1}$, we use $s_{\rm L}=0.239$, $\Lambda_{\rm NL}=325$~MeV, and $C_{\rm SU4}=-4.754\times10^{-6}$~MeV$^{-2}$, with the density smearing defined in Eq.~(\ref{eq:local-smearing}) applied twice to every density operator in the interaction. Both reference interactions are tuned to reproduce the binding energies of medium-mass nuclei to within approximately $20\%$. Table~\ref{tab:Energies_CSU4} compares the ptQMC binding energies of $^4$He and $^{16}$O, decomposed into the nonperturbative reference energies and the first- and second-order perturbative corrections. These calculations use the finer lattice spacing $a=(250~\mathrm{MeV})^{-1}$ to minimize discretization effects. Varying the reference coupling $C_{\rm SU4}$ while keeping the target Hamiltonian fixed substantially redistributes the energy contributions among perturbative orders but produces only small changes in the total binding energies. This stability supports the numerical reliability of the ptQMC predictions at the percent level. Here, perturbation theory serves as a numerical algorithm for solving the target chiral Hamiltonian. It should be distinguished from perturbative renormalization schemes in nuclear EFT, where contributions must be included at specified perturbative orders according to EFT power counting.

\begin{table}[t]
    \renewcommand{\arraystretch}{1.2}
    \setlength{\tabcolsep}{4pt} 
    \centering 
    \caption{LECs and binding energies of selected nuclei from a chiral force built and calibrated at $\Lambda = 400$~MeV as the strength of the nonperturbative SU(4) reference interaction is varied. $E_{0,1,2}$ denote the non-perturbative, first and second order perturbative energies. All energies are in MeV.
    }
    \resizebox{\linewidth}{!}{%
\begin{tabular}{crrrr}   %
        \toprule 
        $a$ (fm) & $0.789$ & $0.789$ & $0.789$ & $0.789$ \\
        \midrule
        $\Lambda$ (MeV) & $400$ & $400$ & $400$ & $400$ \\  
        $C_{\rm SU4}^\prime/C_{\rm SU4}$ & $0.9$ & $1.0$ & $1.1$ & 1.2 \\
        \midrule  
        $E(^4\mathrm{He})$ & $-28.4(2)$ & $-29.0(2)$ & $-28.9(2)$& $-28.7(2)$ \\
        $E_0$ &$-18.1(1)$& $-24.3(1)$ &$-31.1(1)$ & $-38.2(1)$ \\
        $E_1$ &$-1.1(1)$& $4.5(1)$& $11.0(1)$ & $18.1(1)$ \\
        $E_2$ &$-9.2(2)$& $-9.2(2)$ & $-8.8(2)$ & $-8.6(2)$ \\
        \midrule
        $E(^{16}\mathrm{O})$ & $-127.0(9)$ & $-127.0(9)$ & $-127.8(9)$ & $-125.2(10)$  \\
        $E_0$ & $-81.9(1)$ & $-120.8(1)$ & $-159.2(2)$ & $-199.3(2)$ \\
        $E_1$ & $-6.3(1)$ & $32.6(1)$ & $68.5(1)$ & $107.3(1)$ \\
        $E_2$ & $-38.8(9)$ & $-38.8(9)$ & $-37.2(9)$ & $-33.2(10)$  \\
        \bottomrule	 
    \end{tabular}
}
    \label{tab:Energies_CSU4}
\end{table}

The smearing parameters control the range and locality of the
reference interaction. For the spin-paired trial states of even-even
nuclei, the attractive SU(4) interaction provides a nonnegative
auxiliary-field sampling weight~\cite{Lu2022_PRL128-242501}.

\subsection*{Chiral Hamiltonian and density operators}

The contact regulator is separable,
\begin{align}\label{eq:separable-regulator}
f_{\rm abs}(\{\bm p_i',\bm p_i\})
&=\prod_{i=1}^A f_\Lambda(p_i')f_\Lambda(p_i),\nonumber\\
f_\Lambda(p)&=e^{-p^6/(2\Lambda^6)}.
\end{align}
It is implemented through regulated fields
\begin{align}\label{eq:regulated-field}
a_\Lambda(\bm n,\alpha)
&=\frac{1}{L^{3/2}}\sum_{\bm p}e^{i\bm p\cdot\bm n}
f_\Lambda(p)a(\bm p,\alpha)
\nonumber\\
&=\sum_{\bm n'}\widetilde f_\Lambda(\bm n-\bm n')a(\bm n',\alpha),
\end{align}
with $\widetilde f_\Lambda(\bm n)=L^{-3}\sum_{\bm p}
e^{i\bm p\cdot\bm n}f_\Lambda(p)$.
This field regulator is distinct from the Gaussian smearing used
to define $H_0$.

The target Hamiltonian is
\begin{equation}\label{eq:target-Hamiltonian}
H=K+V_{\rm ct}+V_{1\pi}+V_{\rm 3N}+V_{\rm Cou}.
\end{equation}
The contact terms are written using the regulated densities
and currents of Ref.~\cite{Liu2025_EPJA61-85},
\begin{subequations}\label{eq:densities-currents}
\begin{align}
\rho_\Lambda(\bm n,S,I)
&=\sum_{\alpha\beta}a^\dagger_\Lambda(\bm n,\alpha)
[\sigma_S\otimes\tau_I]_{\alpha\beta}a_\Lambda(\bm n,\beta),\\
j_{\Lambda,S,l}(\bm n)
&=\frac{1}{2i}\sum_{\alpha\beta}
\bigl\{a^\dagger_\Lambda(\bm n,\alpha)[\sigma_S]_{\alpha\beta}
\partial_l a_\Lambda(\bm n,\beta)\nonumber\\
&\qquad-[\partial_l a^\dagger_\Lambda(\bm n,\alpha)]
[\sigma_S]_{\alpha\beta}a_\Lambda(\bm n,\beta)\bigr\}.
\end{align}
\end{subequations}
Identity matrices in the isospin space are implicit in the current.
We define $\sigma_0=\tau_0=\mathbb{I}$ and use
$\rho_\Lambda=\rho_\Lambda(\bm n,0,0)$,
$\rho_{\Lambda,S}=\rho_\Lambda(\bm n,S,0)$,
$\rho_{\Lambda,I}=\rho_\Lambda(\bm n,0,I)$, and
$\rho_{\Lambda,SI}=\rho_\Lambda(\bm n,S,I)$.
Spin and Cartesian components take values $1,2,3$, while $I$ labels
isospin components. With the spatial argument suppressed, the
coordinate-space contact interaction is
\begin{align}\label{eq:coordinate-contacts}
V_{\rm ct}&=\frac12\sum_{\bm n}:\!\biggl[
B_1\rho_\Lambda^2+B_2\sum_S\rho_{\Lambda,S}^2
\nonumber\\
&\quad-C_1\rho_\Lambda\nabla^2\rho_\Lambda
-C_2\sum_I\rho_{\Lambda,I}\nabla^2\rho_{\Lambda,I}
\nonumber\\
&\quad-C_3\sum_S\rho_{\Lambda,S}\nabla^2\rho_{\Lambda,S}
-C_4\sum_{SI}\rho_{\Lambda,SI}\nabla^2\rho_{\Lambda,SI}
\nonumber\\
&\quad-\frac{C_5}{2}\sum_{rst}\epsilon_{rst}
\left(\rho_\Lambda\partial_r j_{\Lambda,t,s}
+\rho_{\Lambda,r}\partial_s j_{\Lambda,0,t}\right)
\nonumber\\
&\quad-C_6\sum_{rs}\rho_{\Lambda,r}\partial_r\partial_s\rho_{\Lambda,s}
\nonumber\\
&\quad-C_7\sum_{rsI}\rho_{\Lambda,rI}\partial_r\partial_s\rho_{\Lambda,sI}
\biggr]\!:\, .
\end{align}
Here $r,s,t=1,2,3$, and the spin-density components carry the same
Cartesian labels. Equation~\eqref{eq:coordinate-contacts} follows
the density-current representation in Eq.~(13) of
Ref.~\cite{Liu2025_EPJA61-85}. In particular, both current structures are
retained in the spin-orbit term. Its Fourier transform reproduces
$iC_5(\bm q\times\bm k)\cdot(\bm\sigma_1+\bm\sigma_2)/2$.
The tensor terms contain derivatives of spin densities and therefore
generate momentum transfer rather than average momentum.
Spatial derivatives can be evaluated in momentum space by FFT,
as in Ref.~\cite{Liu2025_EPJA61-85}.

The OPEP uses the unregulated point densities
$\rho(\bm n,S,I)$, obtained by replacing $a_\Lambda$ with $a$.
As in Refs.~\cite{Lu2022_PRL128-242501,Liu2025_EPJA61-85}, only its pion momentum
is regulated in the calculations presented here,
\begin{align}\label{eq:coordinate-OPEP}
V_{1\pi}&=-\frac{g_A^2}{8F_\pi^2}
\sum_{\bm n_1\bm n_2,I}\sum_{S_1S_2}
:\rho(\bm n_1,S_1,I)\nonumber\\
&\quad\times G_\pi(\bm n_1-\bm n_2,S_1,S_2)
\rho(\bm n_2,S_2,I):,
\end{align}
where
\begin{align}\label{eq:pion-propagator}
G_\pi(\bm n,S_1,S_2)
&=\frac{1}{L^3}\sum_{\bm q}e^{i\bm q\cdot\bm n}f_\pi(q^2)
\nonumber\\
&\quad\times\left(\frac{q_{S_1}q_{S_2}}{q^2+M_\pi^2}
+C'_\pi\delta_{S_1S_2}\right).
\end{align}
The limitations of this separate treatment of pion and nucleon modes
are discussed below.

We retain the three-nucleon contact term in the normalization used
in this work,
\begin{equation}\label{eq:three-body-density}
V_{\rm 3N}=-\frac{c_E}{2F_\pi^4\Lambda_\chi}
\sum_{\bm n}\no{\rho_\Lambda^3(\bm n)},
\qquad\Lambda_\chi=700~\mathrm{MeV}.
\end{equation}
The value of $c_E$ is fitted with this normalization at each cutoff.

The static Coulomb interaction follows the coordinate-space
prescription in the Supplemental Material of Ref.~\cite{Lu2022_PRL128-242501},
\begin{equation}\label{eq:Coulomb}
V_{\rm Cou}=\frac{\alpha}{2}\sum_{\bm n_1\bm n_2}
\frac{\no{\rho_p(\bm n_1)\rho_p(\bm n_2)}}
{\max(|\bm n_1-\bm n_2|,1/2)},
\end{equation}
where $\alpha=1/137$ and $\rho_p$ is the point proton density.

\subsection*{Projection and perturbative energy estimators}

We introduce a bookkeeping parameter $\eta$ through
$H(\eta)=H_0+\eta\delta H$, with $\delta H=H-H_0$, and define
\begin{equation}\label{eq:transfer-family}
M(\eta)=\no{e^{-a_t H(\eta)}},\quad
M_0=M(0),\quad
M_1=\left.\frac{\partial M(\eta)}{\partial\eta}\right|_{\eta=0}.
\end{equation}
The temporal spacing is $a_t^{-1}=1000$~MeV, or
$a_t=10^{-3}$~MeV$^{-1}$. The trial state $|\Phi_T\rangle$ is
a Slater determinant of harmonic-oscillator orbitals with a nonzero
overlap with the desired ground state. At a finite half-projection
length $T=L_t/2$, the zeroth-order projected state is
\begin{equation}\label{Psi0}
|\Psi_0(T)\rangle=M_0^T|\Phi_T\rangle.
\end{equation}
Its normalized large-$T$ limit gives the ground state of $H_0$.
The first-order correction is
\begin{equation}\label{Psid1}
|\delta\Psi_1(T)\rangle=
\sum_{k=1}^{T}M_0^{T-k}M_1M_0^{k-1}|\Phi_T\rangle.
\end{equation}
The one-insertion construction of Ref.~\cite{Lu2022_PRL128-242501} uses
$M-M_0$ for $M_1$, with differences beginning at
$\mathcal O(a_t^2)$ in a single time step.
Writing $N_0=\langle\Psi_0|\Psi_0\rangle$ and suppressing the
$T$ arguments, we denote the normalized state by
$|\widehat\Psi(\eta)\rangle=|\Psi(\eta)\rangle/
\sqrt{\langle\Psi(\eta)|\Psi(\eta)\rangle}$. Through first order,
\begin{equation}\label{eq:normalized-state}
\begin{aligned}
|\widehat\Psi(\eta)\rangle
&=\frac{1}{\sqrt{N_0}}\biggl[|\Psi_0\rangle+\eta|\delta\Psi_1\rangle\\
&\quad-\eta\frac{\operatorname{Re}\langle\Psi_0|\delta\Psi_1\rangle}{N_0}
|\Psi_0\rangle\biggr]+\mathcal O(\eta^2).
\end{aligned}
\end{equation}
All projected states in these ratios use the same normalization.
In the small-$a_t$ limit, the energy expansion is
$E(\eta)=E_0+\eta\delta E_1+\eta^2\delta E_2+\mathcal O(\eta^3)$,
with
\begin{subequations}\label{eq:pt-energies}
\begin{align}
E_0&=\frac{\langle\Psi_0|H_0|\Psi_0\rangle}{N_0},\\*
\delta E_1&=\frac{\langle\Psi_0|\delta H|\Psi_0\rangle}{N_0},\\*
\delta E_2&=\frac{\operatorname{Re}
\langle\Psi_0|(\delta H-\delta E_1)|\delta\Psi_1\rangle}{N_0}.
\end{align}
\end{subequations}
The physical result is obtained at $\eta=1$, truncated at the
specified perturbative order. The subtraction in $\delta E_2$
accounts for the normalization correction in
Eq.~\eqref{eq:normalized-state}.

For an inserted operator $O$, the required matrix elements can
be written as
\begin{align}\label{eq:pt-amplitudes}
\mathcal A(O)&=\langle\Phi_T|M_0^T O M_0^T|\Phi_T\rangle,
\nonumber\\
\mathcal A_k(O)&=\langle\Phi_T|M_0^T O
M_0^{T-k}M_1M_0^{k-1}|\Phi_T\rangle.
\end{align}
Consequently, $N_0=\mathcal A(1)$ and
$\langle\Psi_0|O|\delta\Psi_1\rangle=\sum_k\mathcal A_k(O)$.
These amplitudes are evaluated using auxiliary fields sampled with
the SU(4) background weight. The shifted auxiliary-field integration
of Ref.~\cite{Lu2022_PRL128-242501} reduces fluctuations from the perturbing
insertion. The FFT density-current representation above follows
Ref.~\cite{Liu2025_EPJA61-85} and avoids introducing an explicit sum over
excited many-body states as in conventional perturbation theories.

    \begin{figure*}
        \centering
        \includegraphics[width=.85\linewidth]
        {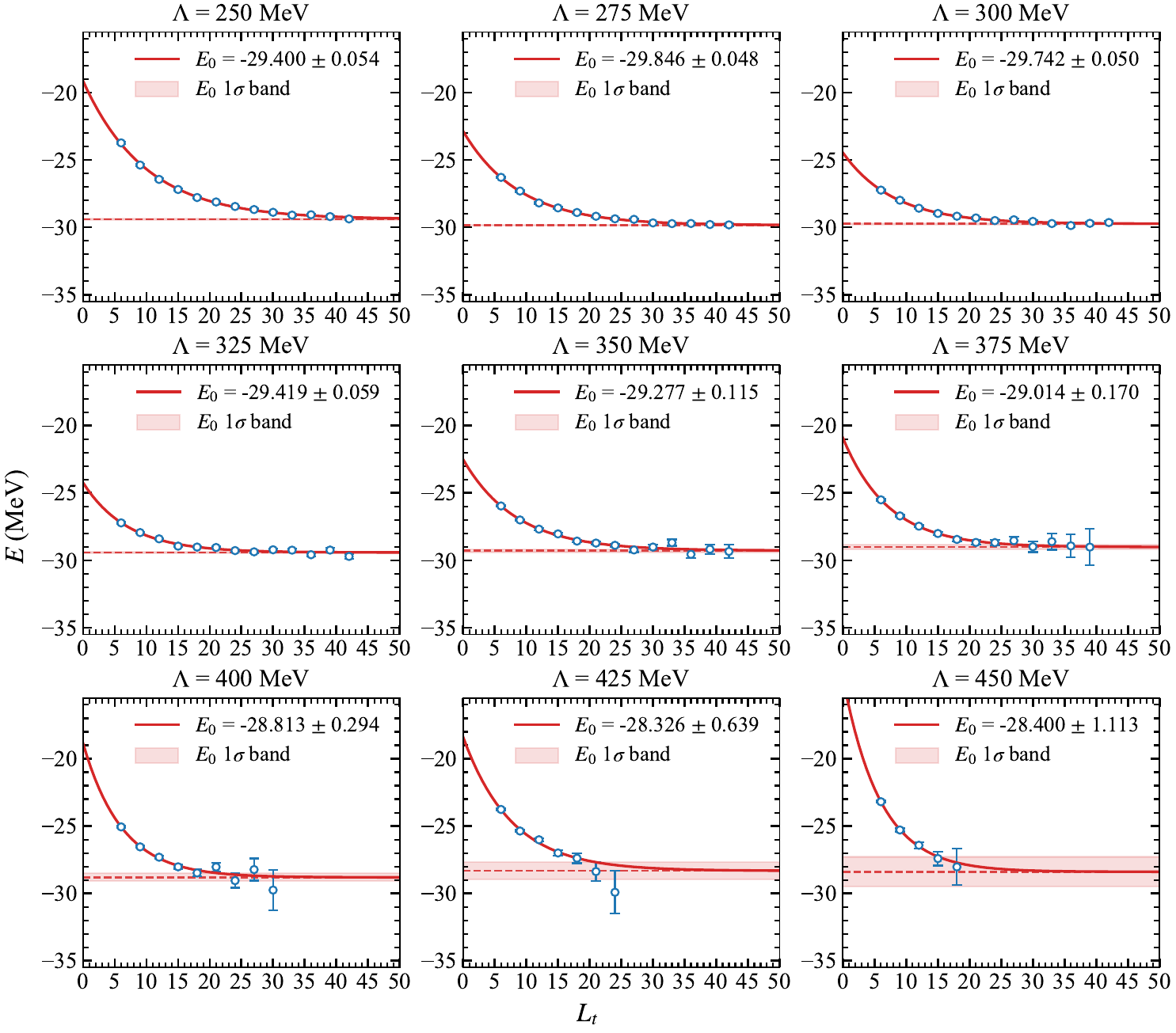}
        \caption{Imaginary-time extrapolation of the \({}^{4}\mathrm{He}\) binding energies obtained from fully non-perturbative calculations including both 2NF and 3NF. The temporal step is \(a_t=10^{-3}\,\mathrm{MeV}^{-1}\). Extrapolated results and uncertainties are marked as red bands.}
        \label{fig:4HeNPT}
    \end{figure*} 

    \begin{figure*}
        \centering
        \includegraphics[width=.85\linewidth]
        {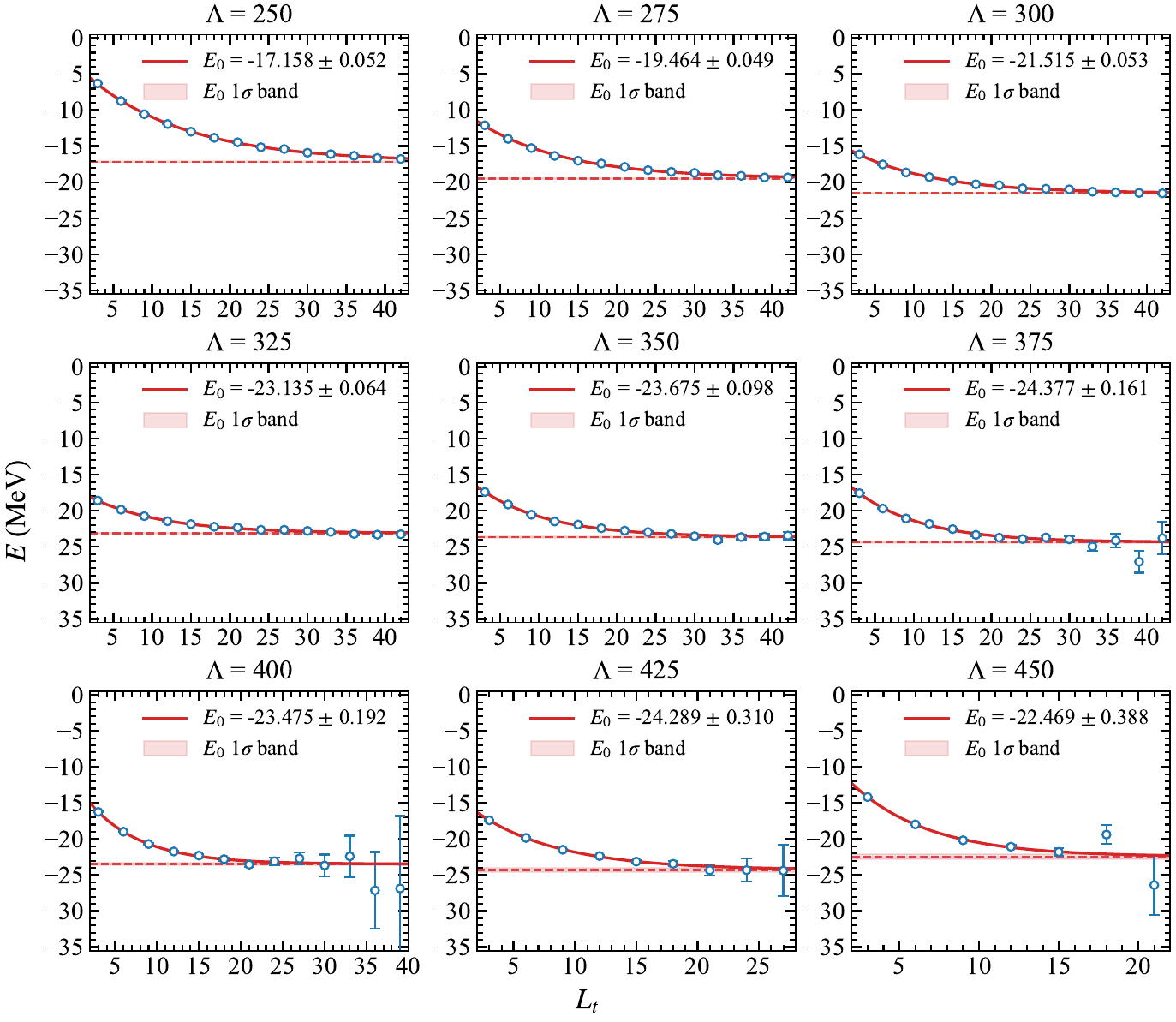}
        \caption{The same as Fig.~\ref{fig:4HeNPT}, but without 3NF.}
        \label{fig:4HeNPT_wo3NF}
    \end{figure*}    

    % To reduce the computational cost, finite-volume effects are
    % examined at $\Lambda=350$~MeV. Boxes with $L=12$ for $A < 30$ and $13$ for $A >= 30$
    % are used to assess the approach to the infinite-volume limit.

    Residual excited-state contributions at finite projection length
    should be fitted with
    \begin{equation}\label{E(Lt)}
    E(L_t)=E_\infty+\mathcal C e^{-a_tL_t\Delta E},
    \end{equation}
    where $\Delta E$ is a fitted energy gap and $E_\infty$ is the
    infinite-time energy. Figures~\ref{fig:4HeNPT}
    and~\ref{fig:4HeNPT_wo3NF} show the fully nonperturbative
    $^4$He benchmarks with and without the three-body force.
    For other nuclei with $A > 4$ we fix the projection time to $\tau = 0.2\mathrm{MeV}^{-1}$.
    Tables~\ref{tab:LECsEnergies} and~\ref{tab:binding} list the
    many-body energies and numerical uncertainties.
    The first ten rows of Tab.~\ref{tab:LECsEnergies} give the LECs.
    The coefficients $B_{1,2}$ and $C_{1\text{--}7}$ are expressed in
    units of $(200~\mathrm{MeV})^{-2}$ and
    $(200~\mathrm{MeV})^{-4}$, respectively.

The calculated binding energy of $^{40}\text{Ca}$ is presented in the main text with two distinct uncertainties. The central value is obtained by averaging the results across the seven cutoff values from $\Lambda=300$ to $450$~MeV. The first uncertainty represents the sample standard deviation over this range, reflecting the cutoff dependence. The second uncertainty is the root-mean-square (RMS) of the combined Monte Carlo and extrapolation errors.

    \begin{table*}[t]
    \renewcommand{\arraystretch}{1.2}
    \setlength{\tabcolsep}{2.4pt} 
    \centering 
    \caption{LECs and binding energies of selected nuclei 
        versus the cutoff variation.
        Rows 1--10 give the LECs in the lattice chiral force 
        with $\Lambda=250\text{--}450$ MeV.
        For each row showing the binding energies of 
        $^3$H, $^{4}$He, $^{16}$O, $^{20}$Ne and $^{40}$Ca, 
        the first line gives the binding energies calculated 
        with $V_{\rm 2NF} + V_{\rm Cou}$, while the second line 
        shows the results obtained from 
        $V_{\rm 2NF} + V_{\rm 3NF} + V_{\rm Cou}$. 
        Numbers in parentheses denote the quoted numerical uncertainties. 
        The LECs $B_{1-2}$ and $C_{1-7}$ are given in units of $(200~\mathrm{MeV})^{-2}$ and $(200~\mathrm{MeV})^{-4}$, respectively. The three-body LEC $c_E$ is dimensionless. For $^4$He, we perform fully nonperturbative Monte Carlo calculations with imaginary-time extrapolation. For nuclei with $A>4$, we use ptQMC at a fixed projection length $L_t=200$. All energies are in MeV.}
 \resizebox{\linewidth}{!}{%
\begin{tabular}{crrrrrrrrrr}
    \toprule 
    $\Lambda$ (MeV)&$250$&$275$&$300$&$325$&$350$
                   &$375$&$400$&425&450&Exp.\\  
    \midrule  
	$B_1$ &$-$4.931&$-$4.762&$-$4.592&$-$4.435&$-$4.285&$-$4.140&$-$3.997&$-$3.852&$-$3.642\\
    $B_2$ &$-$0.369&$-$0.326&$-$0.285&$-$0.246&$-$0.206&$-$0.164&$-$0.119&$-$0.072&$-$0.010\\
    $C_1$ &0.363&0.432&0.454&0.456&0.449&0.440&0.432&0.427&0.429\\
    $C_2$ &0.063&0.011&$-$0.017&$-$0.032&$-$0.041&$-$0.048&$-$0.056&$-$0.064&$-$0.077\\
    $C_3$ &$-$0.002&$-$0.024&$-$0.033&$-$0.039&$-$0.042&$-$0.047&$-$0.051&$-$0.056&$-$0.063\\
    $C_4$ &0.008&$-$0.029&$-$0.049&$-$0.060&$-$0.067&$-$0.073&$-$0.078&$-$0.086&$-$0.094\\
    $C_5$ &0.997&0.939&0.901&0.875&0.856& 0.841&0.829&0.819&0.814\\
    $C_6$ & 0.024&0.015&0.007&0.002&$-$0.002&$-$0.004&$-$0.005&$-$0.008&$-$0.005\\
    $C_7$ &$-$0.288&$-$0.267&$-$0.255&$-$0.248&$-$0.245&$-$0.245&$-$0.247&$-$0.250&$-$0.260\\   
    $c_E$&5.170&2.763&1.538&0.890&0.561&0.412&0.380 &0.410&0.589\\
        \midrule  
   $E(^3$H)$^\mathrm{w/o}_\mathrm{3NF}$

        &$-$6.174&$-$6.634&$-$7.052&$-$7.392&$-$7.641
    &$-$7.771&$-$7.781&$-$7.750&$-$7.590& \\
    $E(^3$H)&$-$8.482&$-$8.482&$-$8.482&$-$8.482&$-$8.482&$-$8.482&$-$8.482&$-$8.482&$-$8.482&$-$8.482 \\
    \midrule  E($^4$He)$^\mathrm{w/o}_\mathrm{3NF}$&$-$17.2(1)&$-$19.5(1)&$-$21.5(1)&$-$23.1(1)&$-$23.7(1)&$-$24.4(2)&$-$23.5(2)&$-$24.3(3)&$-$22.4(4)\\
    $E(^4$He)&$-$29.4(1)&$-$29.8(1)&$-$29.7(1)&$-$29.4(1)&$-$29.3(1)&$-$29.0(2)&$-$28.8(3)&$-$28.3(6)&$-$28.4(11)\\
  
    \midrule
    $E(^{16}$O)$^\mathrm{w/o}_\mathrm{3NF}$ 
    &$-$77.1(7)&$-$78.8(8)&$-$85.4(7)&$-$93.8(7)&$-$100.8(7)
    &$-$104.3(7)&$-$104.1(8)&$-$100.3(8)&$-$96.4(10)&  \\
    $E(^{16}$O)&$-$135.8(20)&$-$124.8(14)&$-$124.5(11)&$-$126.3(9)&$-$127.3(8)
    &$-$128.1(8)&$-$128.1(8)&$-$127.7(9)&$-$129.3(11)&$-$127.6 \\ 
    \midrule
    $E(^{20}$Ne)$^\mathrm{w/o}_\mathrm{3NF}$ 
    &$-$100.8(16)&$-$101.9(14)&$-$107.7(13)&$-$116.7(13)
    &$-$124.6(13)&$-$128.4(13)&$-$128.0(15)&$-$126.9(16)&$-$121.6(20)&\\
    $E(^{20}$Ne)&$-$178.7(30)&$-$162.5(21)&$-$157.8(16)&$-$159.6(14)
    &$-$160.9(12)&$-$160.8(12)&$-$161.1(15)&$-$160.2(15)&$-$160.7(17)&$-$160.6 \\
    \midrule
    $E(^{40}$Ca)$^\mathrm{w/o}_\mathrm{3NF}$
    &$-$180.3(24)&$-$204.5(19)&$-$224.6(17)&$-$247.6(17)&$-$266.8(17)
    &$-$274.3(17)&$-$266.1(18)&$-$255.5(22)&$-$217.5(28)&\\     
    $E(^{40}$Ca)&$-$391.6(62)&$-$355.8(40)&$-$346.5(28)&$-$341.6(24)&$-$346.4(15)
    &$-$341.3(20)&$-$344.7(20)&$-$346.2(24)&$-$340.5(39)&$-$342.1 \\  

    \bottomrule	 
    \end{tabular}
}
\label{tab:LECsEnergies}
\end{table*}

\begin{figure*}[t]
    \centering
        \includegraphics[width=1\linewidth]{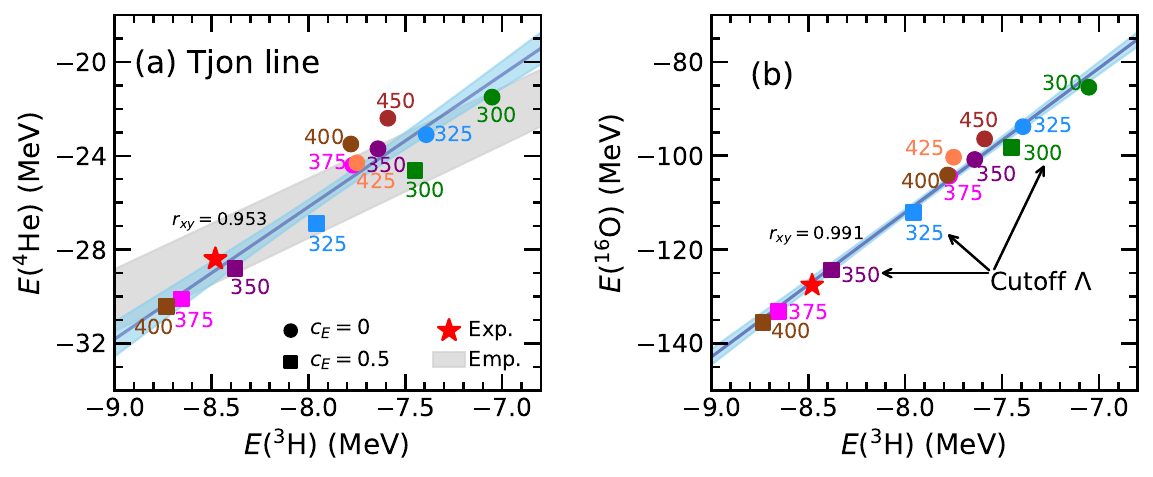}
        \caption{(a) The Tjon line (band) of $^3$H and $^4$He
        binding energies calculated by varying cutoffs 
        over the values indicated next to the points, for 2NF only and
        2NF + 3NF with $c_E=0.5$.
        The empirical Tjon band is indicated by 
        shaded area, adopted from Ref.~\cite{Platter2005_PLB607-254}.
        (b) Binding energy correlation between 
        $^3$H and $^{16}$O with the same notation as in panel~(a). }  
        \label{fig:H3He4O16corr}
    \end{figure*}

    We also examine the Tjon correlation between the triton and
    alpha-particle energies~\cite{tjon1975plb}. Figure~\ref{fig:H3He4O16corr}(a)
    compares calculations with 2NF only, $c_E=0$, and calculations
    with $c_E=0.5$ held fixed while the cutoff is varied.
    The results follow the empirical Tjon band~\cite{Platter2005_PLB607-254}.
    For the points displayed in the figure, the Pearson coefficient
    is $r_{xy}=0.953$. Galilean-invariance restoration improves the
    few-body correlation and its relation to the experimental point,
    as shown in Ref.~\cite{Shi2026_PLB874-140303}.
    Figure~\ref{fig:H3He4O16corr}(b) shows the analogous correlation
    between the triton and $^{16}$O energies, with $r_{xy}=0.991$.
    These correlations support the use of a common three-body coupling
    calibrated to the triton when predicting the heavier systems.

    The observed correlations suggest that the information
    supplied by additional three-body couplings should be assessed
    together with the regularization prescription. Establishing the
    extent of this behavior requires further calculations.

\section*{Brueckner-Hartree-Fock approach}
    Nuclear matter with a sign-problematic Hamiltonian is
    challenging for lattice Monte Carlo. We therefore employ the
    Brueckner-Hartree-Fock (BHF) approach. Three-nucleon forces are
    reduced to density-dependent two-body interactions by normal
    ordering~\cite{hebeler2021pr}.
    Particle-particle ladders are resummed by the Bethe-Goldstone
    equation~\cite{Day1967_RMP39-719},
    \begin{equation}\label{Beq}
    G(W)=V+V\frac{Q}{W-\epsilon_1-\epsilon_2+i0}G(W).
    \end{equation}
    At zero temperature the Pauli operator is
    \begin{equation}
    Q(\bm p_1,\bm p_2)=
    \theta(|\bm p_1|-p_F)\theta(|\bm p_2|-p_F),
    \end{equation}
    with $p_F=(3\pi^2\rho/2)^{1/3}$ for SNM of total density $\rho$.
    The denominator acts on an intermediate state as
    \begin{equation}
    e(W)|\bm p_1\bm p_2\rangle
    =[W-\epsilon(p_1)-\epsilon(p_2)+i0]|\bm p_1\bm p_2\rangle.
    \end{equation}
    We use the continuous choice~\cite{Song1998_PRL81-1584},
    \begin{align}\label{eq:BHF-spectrum}
    \epsilon(p)&=\frac{p^2}{2M_{\rm N}}+\operatorname{Re}U(p),
    \nonumber\\
    U(p)&=\sum_{\sigma'\tau'}\int_{p'<p_F}\frac{d^3p'}{(2\pi)^3}
    \langle\bm p\bm p'|G(W)|\bm p\bm p'\rangle_{\mathcal A},
    \end{align}
    where $W=\epsilon(p)+\epsilon(p')$. Spin-isospin labels on the
    external state are implicit, and $\mathcal A$ denotes
    antisymmetrization. The two-body-like energy expression is
    \begin{align}\label{eq:BHF-energy}
    \frac{E(\rho)}{A}&=\frac{3p_F^2}{10M_{\rm N}}+
    \frac{1}{2\rho}\sum_{\sigma_1\tau_1\sigma_2\tau_2}
    \int_{p_1,p_2<p_F}\frac{d^3p_1d^3p_2}{(2\pi)^6}
    \nonumber\\
    &\quad\times\operatorname{Re}
    \langle\bm p_1\bm p_2|G(\epsilon_1+\epsilon_2)
    |\bm p_1\bm p_2\rangle_{\mathcal A}.
    \end{align}

    We use an angular-averaging approximation for the dependence
    of the regulator on the pair center-of-mass momentum. Defining
    $P=|\bm P_{\rm cm}|$, $x=\widehat{\bm P}_{\rm cm}\cdot\widehat{\bm p}$,
    and $A_p=p^2+P^2/4$, the averaged pair factor is
    \begin{align}\label{eq:angle-average}
    \bar f(P,p)&=\frac12\int_{-1}^{1}dx\,
    \exp\left[-\frac{A_p^3+3A_pP^2p^2x^2}{\Lambda^6}\right]
    \nonumber\\
    &=e^{-A_p^3/\Lambda^6}\frac{\sqrt\pi}{2s_p}\operatorname{erf}(s_p),
    \end{align}
    where $s_p=Pp\sqrt{3A_p}/\Lambda^3$ and the continuous limit is
    taken when $s_p=0$. The two-body regulator is approximated by
    \begin{equation}\label{fabs:AA}
    f_{\rm abs}\approx\bar f(P,p')\bar f(P,p).
    \end{equation}
    This prescription averages the regulator factors rather than
    performing an exact projection of their product with every
    interaction operator.
    The normal-ordered 3NF is obtained by 
	\begin{align} \label{NO3N}
		\langle 1'2'|V_{\overline{\mathrm{3N}}}
        |12\rangle_\mathcal{A} 
		= \sum_{\sigma_3 \tau_3} \int_{|\bm p_3|<p_F}\frac{d^3p_3}{(2\pi)^3} 
		\langle 1'2'3|V_\mathrm{3N}|123\rangle_\mathcal{A},
	\end{align} 
    where $|i\rangle =|\bm{p}_i\sigma_i\tau_i\rangle$,  
    $\sigma_i$, $\tau_i$ are the spin and isospin indices and
    the subscript $\mathcal{A}$ indicates antisymmetrization.

    We use $V=V_\mathrm{2N} + V_{\overline{\mathrm{3N}}}/3$ 
    as the kernel in the Bethe-Goldstone equation~\eqref{Beq}.
    The factor $1/3$ reproduces the three-body counting in the HF
    energy. Equations~\eqref{eq:BHF-spectrum} and~\eqref{eq:BHF-energy}
    specify a density-dependent two-body prescription, rather than a
    complete treatment of three-body rearrangement contributions.
    In the HF single-particle field, the corresponding factor is
    $1/2$ instead of $1/3$~\cite{Kohno2013_PRC88-064005}.

    For the separable contact interaction, the normal-ordered
    two-body contribution has the following form in the
    ${}^1S_0$ and ${}^3S_1$ channels after the angular averaging
    \begin{equation}\label{NOV3N}
        \langle S|V_{\overline{\mathrm{3N}}}|S\rangle  
        =-\frac{3c_E \rho_\Lambda}{4F_\pi^4 \Lambda_\chi} ~
        f_{\rm abs}(p_1',p_1,p_2',p_2),
    \end{equation}   
    where $\rho_\Lambda$ is the regulated total nucleon density,
    \begin{align}\label{eq:regulated-density}
    \rho_\Lambda&=\sum_{\sigma\tau}\int_0^{p_F}
    \frac{p^2dp}{2\pi^2}e^{-(p/\Lambda)^6}
    \nonumber\\
    &=\frac{2\Lambda^3}{3\pi^{3/2}}
    \operatorname{erf}\left(\frac{p_F^3}{\Lambda^3}\right).
    \end{align}
    The fourfold spin-isospin degeneracy ensures
    $\rho_\Lambda\to\rho$ as $\Lambda\to\infty$.
    With the normalization in Eq.~\eqref{eq:three-body-density},
    the unregulated HF limit is
    $E_{3N}^{\rm HF}/A=-3c_E\rho^2/(16F_\pi^4\Lambda_\chi)$.

    \begin{figure} 
        \centering
        \includegraphics[width=1.0\columnwidth]{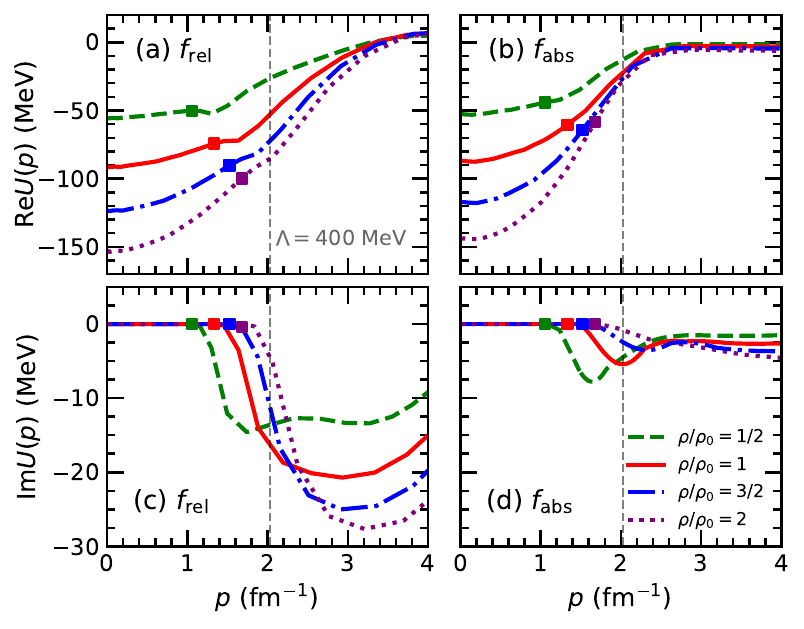}
        \caption{Real and imaginary parts of single-particle potentials
        in symmetric nuclear matter at various densities calculated using 2NF with cutoff $\Lambda=400$~MeV.
        Results with the same LECs and 
        relative-momentum regulator are contrasted.
        Squares denote the corresponding Fermi momenta.} 
        \label{fig:snm_usp}
    \end{figure}

    At a fixed density, the kinetic-energy contribution is the same
    for both regulator prescriptions. Their EOS differences therefore
    arise from the interaction energy.
    We compare in Fig.~\ref{fig:snm_usp} the single-particle potentials $U(p)$ generated by both regulators at $\Lambda = 400\;\text{MeV}$.
    While both regulators yield similar results for $p \ll \Lambda$, $f_{\mathrm{abs}}$ suppresses high-momentum attraction far more efficiently.
    In contrast, $f_{\mathrm{rel}}$ exhibits a much slower high-momentum damping, leading to the excessive binding.
    The distinction becomes even more pronounced in the imaginary part, $\text{Im}\,U(p)$, which emerges for $p > p_F$ and reflects scattering and damping of particles above the Fermi surface.
    This marked contrast highlights the ability of the \emph{absolute}-momentum regulator to suppress high-momentum contributions, thereby reducing the excessive attraction in the nuclear-matter calculations.

\begin{figure} 
        \centering
        \includegraphics[width=1.0\columnwidth]{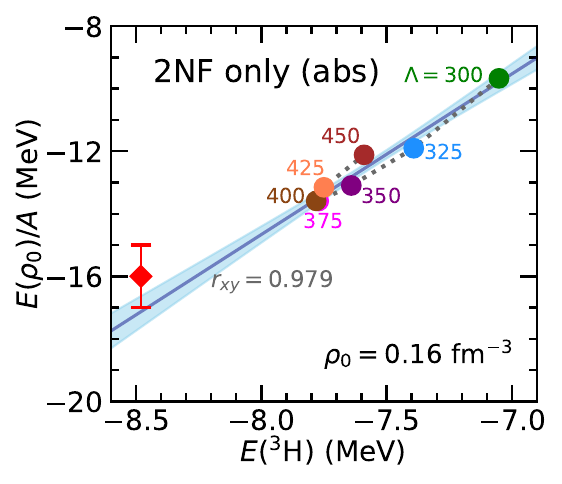}
        \caption{The binding energy per nucleon versus triton binding energy calculated using 2NF regulated according to absolute momentum.} 
        \label{fig:eosh3}
    \end{figure}

    Figure~\ref{fig:eosh3} reveals a tight linear correlation between the 2NF-only values of $E(\rho_0)/A$ and the triton energy obtained with $f_{\rm abs}$. Analogous to the familiar Tjon and Coester correlations~\cite{Coester1970_PRC1-769}, this relation supports the view that the universality associated with the anomalously large NN scattering lengths connects binding from few-nucleon systems to bulk nuclear matter.

\section{Dependence on OPEP cutoff $\Lambda_\pi$}

In the calculations presented in the main text, the cutoff of the one-pion-exchange potential (OPEP) is fixed at $\Lambda_\pi = 300$~MeV. Here, we examine the sensitivity of our results to this choice. For simplicity, we fix the contact-interaction cutoff at $\Lambda = 400$~MeV and vary $\Lambda_\pi$ from $300$ to $400$~MeV in steps of $50$~MeV. At each value of $\Lambda_\pi$, we independently refit the LECs to the NN phase shifts and the triton binding energy. The resulting LECs are listed in Table~\ref{tab:LambdaPiDependence}.

To quantify the effect of OPEP at each cutoff, we also calculate the energies with OPEP omitted while keeping all other interaction terms fixed. These energies are denoted by $E(^4\mathrm{He})^\mathrm{w/o}_\mathrm{OPEP}$ and $E(^{16}\mathrm{O})^\mathrm{w/o}_\mathrm{OPEP}$. Their differences from the full results measure the net energy shifts induced by OPEP. As $\Lambda_\pi$ increases from $300$ to $400$~MeV, these shifts change from approximately $-10$ to $-17$~MeV for $^4$He and from $-40$ to $-76$~MeV for $^{16}$O. The increasing attraction reflects the growing importance of higher-momentum components of OPEP as its cutoff is raised. Nevertheless, the refitted contact interactions provide compensating contributions, yielding stable total energies throughout this cutoff interval.

\begin{figure}
        \centering  
        \includegraphics[width=0.9\columnwidth] {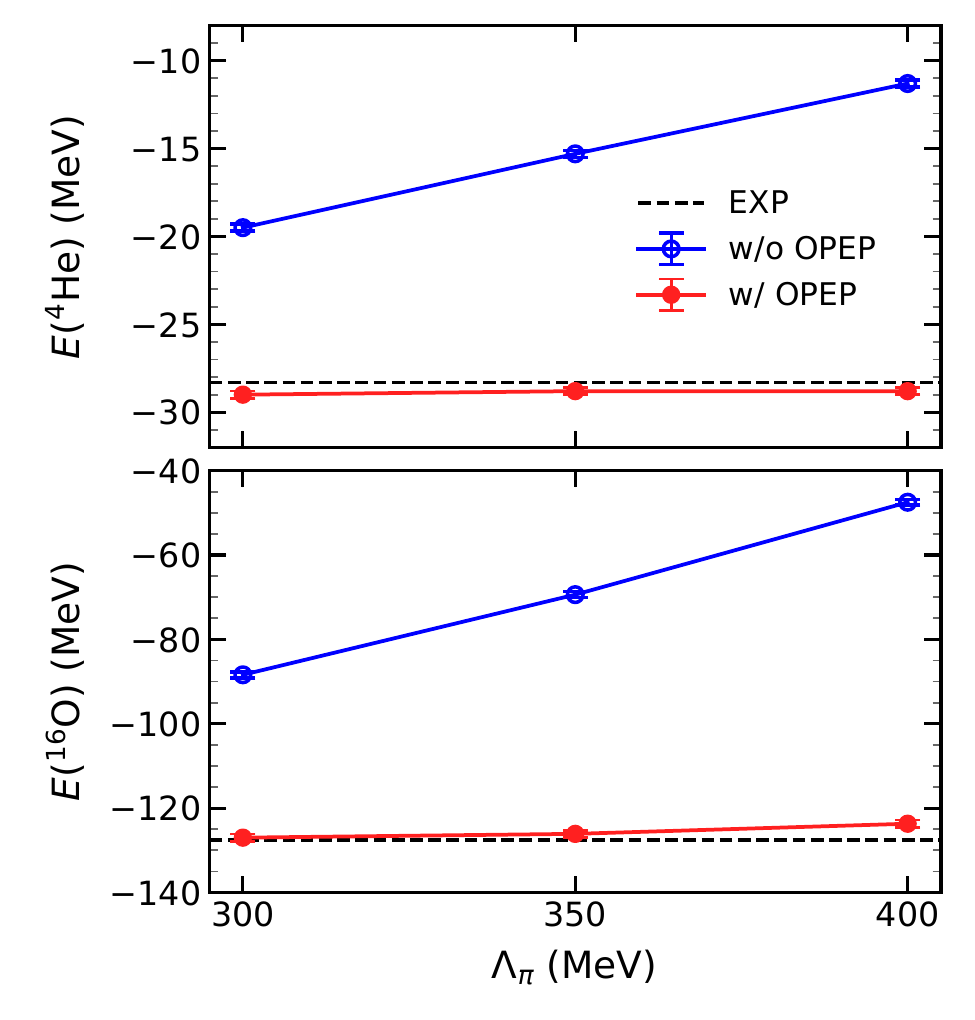}
        \caption{Binding energies calculated with and without OPEP. For each value of the OPEP cutoff \(\Lambda_\pi\), the contact-term LECs were refitted to the NN phase shifts and the triton binding energy.}
        \label{fig:BindingEnergiesOPEP}
    \end{figure}

To examine this stability further, we decompose the total energies into their zeroth-, first-, and second-order perturbative contributions, denoted by $E_0$, $E_1$, and $E_2$, respectively. For calculations with OPEP included, $E_0$ remains nearly constant because the same nonperturbative SU(4) Hamiltonian is used throughout. By contrast, the individual first- and second-order contributions vary
substantially over the range of $\Lambda_\pi$ considered. Their changes
largely compensate, leaving the total energies comparatively stable. This cancellation is associated with SU(4) symmetry breaking, as demonstrated in Ref.~\cite{Lu2022_PRL128-242501}. These results are also shown in Fig.~\ref{fig:BindingEnergiesOPEP}.

We note an inconsistency in our regularization scheme. A consistent implementation of Wilsonian RG requires the nucleon fields entering pion--nucleon vertices to be regulated with the same cutoff as those entering NN contact vertices. Integrating out the pion fields then introduces additional nonlocal single-particle regulator factors into the pion-exchange potentials.

Our results nevertheless suggest that including these factors would preserve the main conclusions of this work. In the present calculations, the contributions from the 2NF and 3NF contact terms vary with $\Lambda$, but these variations compensate one another, leaving the predicted total energies nearly invariant. Meanwhile, the OPEP contribution remains nearly constant. Regulating the nucleon fields at pion--nucleon vertices would introduce an additional $\Lambda$ dependence into OPEP. The resulting changes would primarily affect its high-momentum components and are therefore expected to be absorbed by corresponding adjustments of the contact terms, analogous to the compensation observed when varying $\Lambda_\pi$.

\begin{table}[t]
    \renewcommand{\arraystretch}{1.2}
    \setlength{\tabcolsep}{6pt}
    \centering
    \caption{LECs and binding energies of selected nuclei
        as functions of $\Lambda_\pi$, with $g_A=1.287$, spatial
        lattice spacing $a=0.789$~fm, and contact cutoff $\Lambda=400$~MeV.
        The units of $B_{1,2}$ and $C_{1\text{--}7}$ are
        $(200~\mathrm{MeV})^{-2}$ and $(200~\mathrm{MeV})^{-4}$,
        respectively, and $c_E$ is dimensionless. Energies are in MeV.
    }
    \resizebox{\linewidth}{!}{%
\begin{tabular}{crrrr}   %
        \toprule 
        $\Lambda_\pi$ (MeV) & $300$ & $350$ & $400$ & $\mathrm{Exp.}$\\
        \midrule  
        $B_1$ & $-4.006$ & $-3.634$ & $-3.296$ & \\
        $B_2$ & $-0.121$ & $-0.036$ & $0.045$ & \\
        $C_1$ & $0.431$ & $0.414$ & $0.395$ & \\
        $C_2$ & $-0.054$ & $-0.032$ & $-0.015$ & \\
        $C_3$ & $-0.050$ & $-0.080$ & $-0.102$ & \\
        $C_4$ & $-0.080$ & $-0.076$ & $-0.078$ & \\
        $C_5$ & $0.828$ & $0.805$ & $0.793$ & \\
        $C_6$ & $-0.008$ & $0.052$ & $0.093$ & \\
        $C_7$ & $-0.245$ & $-0.182$ & $-0.127$ & \\   
        $c_E$ & $0.350$ & $0.357$ & $0.349$ & \\
       \midrule  
        $E_\infty(^3\mathrm{H})$ & $-8.482$ & $-8.482$ & $-8.482$ & $-8.482$\\
        \midrule
        $E(^4\mathrm{He})^\mathrm{w/o}_\mathrm{OPEP}$ & $-19.5(2)$ & $-15.3(2)$ & $-11.3(2)$ & \\ 
        $E_0$ & $-24.3(1)$ & $-24.3(1)$ & $-24.3(1)$ & \\
        $E_1$ & $9.1(1)$ & $11.9(1)$ & $15.2(1)$ & \\
        $E_2$ & $-4.2(2)$ & $-2.9(2)$ & $-2.2(2)$ & \\
        \midrule
        $E(^4\mathrm{He})$ & $-29.0(2)$ & $-28.8(2)$ & $-28.8(2)$ & $-28.3$ \\ 
        $E_0$ & $-24.3(1)$ & $-24.3(1)$ & $-24.3(1)$ & \\ 
        $E_1$ & $4.5(0)$ & $5.9(1)$ & $7.8(1)$ & \\ 
        $E_2$ & $-9.2(2)$ & $-10.3(2)$ & $-12.2(2)$ & \\ 
        \midrule
        $E(^{16}\mathrm{O})^\mathrm{w/o}_\mathrm{OPEP}$ & $-88.4(7)$ & $-69.4(7)$ & $-47.5(7)$ & \\ 
        $E_0$ & $-120.9(1)$ & $-120.8(1)$ & $-120.8(1)$ & \\ 
        $E_1$ & $51.4(1)$ & $65.8(1)$ & $83.4(1)$ & \\ 
        $E_2$ & $-18.9(7)$ & $-14.3(7)$ & $-10.2(7)$ & \\ 
        \midrule
        $E(^{16}\mathrm{O})$ & $-127.0(9)$ & $-126.1(8)$ & $-123.7(9)$ & $-127.6$\\ 
        $E_0$ & $-120.8(1)$ & $-120.9(1)$ & $-120.8(1)$ & \\ 
        $E_1$ & $32.6(1)$ & $40.1(1)$ & $51.4(1)$ & \\ 
        $E_2$ & $-38.8(8)$ & $-45.3(8)$ & $-54.3(9)$ & \\ 
        \bottomrule	 
    \end{tabular}
}
    \label{tab:LambdaPiDependence}

\end{table}

\section{Galilean invariance restoration}

\begin{figure}[htbp] 
\centering
\includegraphics[width=1.0\columnwidth]{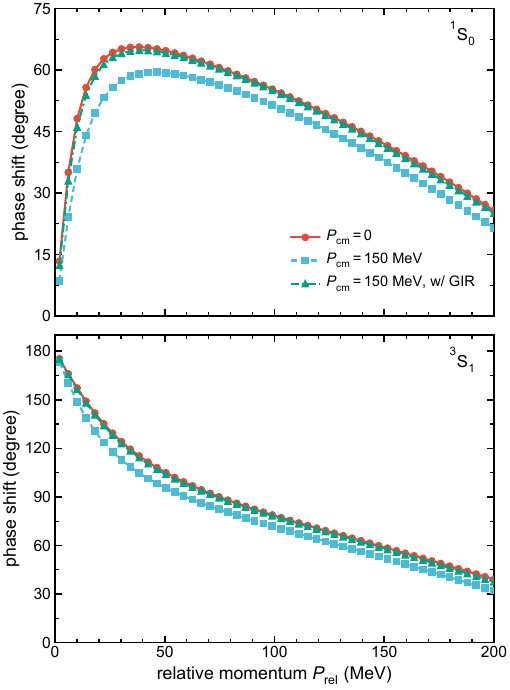}
\caption{NN S-wave phase shifts calculated in the center-of-mass frame \(P_\mathrm{cm}=0\) (red dots) and in a moving frame with \(P_\mathrm{cm}=150~\mathrm{MeV}\) without (cyan squares) and with (green triangles) Galilean invariance restoration terms. OPEP is included, $\Lambda = \Lambda_\pi = 300$~MeV. $S$-$D$ mixing is neglected.}
\label{fig:SwaveGIR}
\end{figure} 

\begin{figure}[htbp] 
\centering
\includegraphics[width=1.0\columnwidth]{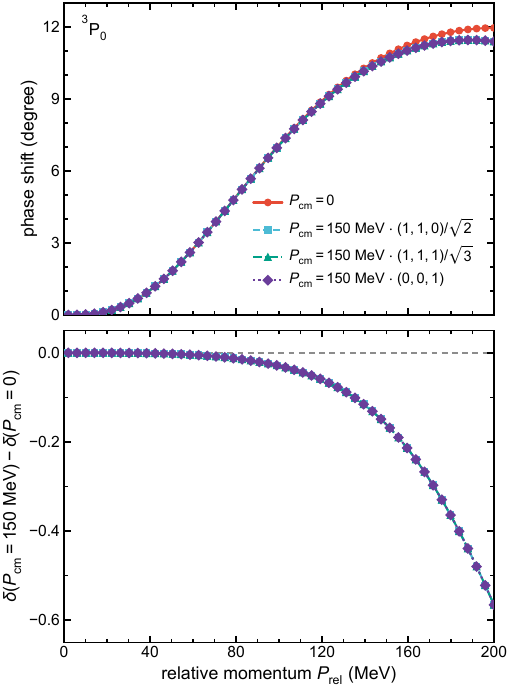}
\caption{Upper panel, NN $^3$P$_0$ phase shifts calculated in the center-of-mass frame (\(P_\mathrm{cm}=0\)) (red dots) and in moving frames with \(|\mathbf{P}_\mathrm{cm}| = 150~\mathrm{MeV}\) along different directions. Lower panel, the difference between  $P_\mathrm{cm}=150$~MeV and $P_\mathrm{cm}=0$ results. OPEP is included, $\Lambda = \Lambda_\pi = 300$~MeV.}
\label{fig:PwaveGIR}
\end{figure} 

\begin{figure}[htbp] 
\centering
\includegraphics[width=1.0\columnwidth]{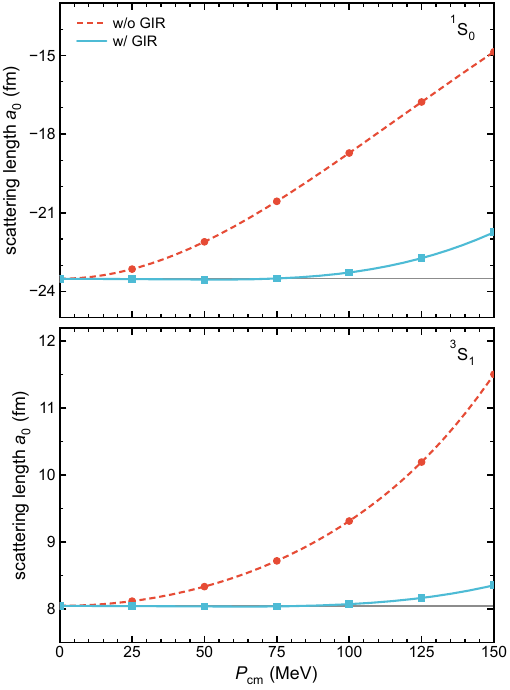}
\caption{Scattering lengths of the \(^{1}S_{0}\) and \(^{3}S_{1}\) partial waves for NN scattering as functions of the center-of-mass momentum \(P_{\mathrm{cm}}\). The dashed and solid lines represent the results without and with the Galilean invariance restoration terms, respectively. OPEP and $S$--$D$ mixing are omitted in this diagnostic, so
the center-of-mass scattering lengths need not equal their physical values.
The contact LECs are taken from the fit with
$\Lambda=\Lambda_\pi=300$~MeV.}
\label{fig:011}
\end{figure}

In this work, we regulate the \emph{absolute} single-particle momenta, thereby explicitly breaking Galilean invariance. Such effects are familiar in lattice calculations, where the lattice selects a preferred inertial frame. 

To illustrate the origin of this breaking, we compare \emph{absolute}- and \emph{relative}-momentum regulators defined with the same cutoff $\Lambda$. For a soft Gaussian regulator, the corresponding two-body regulator factors are
\begin{align}
    f_{\text{rel}}  &= \exp\left(-\frac{p^2 + p^{\prime2}}{\Lambda^2}\right), \\
    f_{\text{abs}}  &= \exp\left(-\frac{p_1^2 + p_2^2 + p_1^{\prime2} + p_2^{\prime2}}{2\Lambda^2 } \right) \nonumber \\
    &= \exp\left(-\frac{P_{\text{cm}}^2}{2\Lambda^2}\right) \exp\left(-\frac{p^2 + p^{\prime2}}{\Lambda^2}\right),
\end{align}
where $\bm{p} = (\bm{p}_1 - \bm{p}_2)/2$ and $\bm{p}^\prime = (\bm{p}_1^\prime - \bm{p}_2^\prime)/2$ are the incoming and outgoing relative momenta, respectively, and momentum conservation implies $\bm{P}_{\text{cm}} = \bm{p}_1 + \bm{p}_2 = \bm{p}_1^\prime + \bm{p}_2^\prime$. In the two-body center-of-mass frame, $\bm{P}_{\text{cm}}=\bm{0}$, the two regulators coincide and yield the same fitted two-body LECs.

Away from this frame, $f_{\text{abs}}$ contains an additional factor that suppresses the interaction for pairs with large center-of-mass momenta. For predominantly attractive interactions, this suppression tends to weaken many-body binding relative to the \emph{relative}-momentum prescription. Thus, reproducing identical two-body observables in the center-of-mass frame does not by itself guarantee identical many-body predictions for the two regularization schemes.

\begin{table}[htbp]
    \renewcommand{\arraystretch}{1.2}
    \setlength{\tabcolsep}{6pt}
\centering
\caption{Binding energies (MeV) for $N=Z$ nuclei calculated at $\Lambda = 350$~MeV. Numbers in parentheses denote the uncertainty in the last digit, e.g., $-126.1(6)$ means $-126.1 \pm 0.6$~MeV. The last column gives the ratio of the 2NF-only result to the 2NF+3NF result in percent.}
\label{tab:binding}
\resizebox{\linewidth}{!}{%
\begin{tabular}{l c c c c}
\toprule
Nucleus & 2NF only & 2NF+3NF & EXP & Ratio [\%] \\
\midrule
$^4$He   & $-23.7(4)$  & $-29.0(2)$  & $-28.3$  & 81.7 \\
$^{16}$O & $-100.8(7)$ & $-126.1(6)$ & $-127.6$ & 79.94 \\
$^{20}$Ne & $-124.6(13)$ & $-159.0(16)$ & $-160.6$ & 78.36 \\
$^{24}$Mg & $-157.7(11)$ & $-189.3(44)$ & $-198.3$ & 83.31 \\
$^{28}$Si & $-186.7(14)$ & $-217.3(10)$ & $-236.5$ & 85.92 \\
$^{32}$S  & $-208.9(15)$ & $-260.9(22)$ & $-271.8$ & 80.07 \\
$^{36}$Ar & $-243.6(43)$ & $-294.9(19)$ & $-306.7$ & 82.60 \\
$^{40}$Ca & $-266.8(17)$ & $-340.6(27)$ & $-342.1$ & 78.33 \\
\bottomrule
\end{tabular}
}
\end{table}

Galilean-invariance-breaking effects and their correction have been studied extensively within nuclear lattice effective field theory (NLEFT). In a one-dimensional toy model, Galilean invariance restoration (GIR) terms constructed following Symanzik's improvement program were shown to accelerate the convergence of lattice predictions~\cite{klein2018epja}. GIR terms have also been incorporated into calculations with realistic nuclear forces to improve agreement with experimental data \cite{Li2019_PRC99-064001,Elhatisari2024_Nature630-59}.

\begin{table*} 
    \renewcommand{\arraystretch}{1.2}
    \setlength{\tabcolsep}{2.0pt} 
    \centering 
    \caption{Low-energy constants with Galilean invariance restoration terms.
    The LECs $B_{1-2}$ and $C_{1-7}$ are given in units of $(200~\mathrm{MeV})^{-2}$ and $(200~\mathrm{MeV})^{-4}$, respectively. 
    The $g_{{}^1S_0/{}^3S_1}$ and $g'_{{}^1S_0/{}^3S_1}$
    coefficients are given in units of $(200~\mathrm{MeV})^{-4}$
    and $(200~\mathrm{MeV})^{-6}$, respectively.  The dimensionless 
    three-body LEC $c_E$ is fitted to the triton binding energy E($^3$H)=$-$8.482 MeV.
    The two-body contact terms and OPEP remain the same as in the previous section.
    The OPEP cutoff is fixed to $\Lambda_\pi= 300$~MeV.
    All energies are in MeV.}
\resizebox{\linewidth}{!}{%
\begin{tabular}{crrrrrrrrrr}   
\toprule 
$\Lambda$ (MeV) & 250 & 275 & 300 & 325 & 350 & 375 & 400 & 425 & 450 & Exp. \\
\midrule
$B_1$ & $-4.931$ & $-4.762$ & $-4.592$ & $-4.435$ & $-4.285$ & $-4.140$ & $-3.997$ & $-3.852$ & $-3.642$ & \\
$B_2$ & $-0.369$ & $-0.326$ & $-0.285$ & $-0.246$ & $-0.206$ & $-0.164$ & $-0.119$ & $-0.072$ & $-0.010$ & \\
$C_1$ & 0.363 & 0.432 & 0.454 & 0.456 & 0.449 & 0.440 & 0.432 & 0.427 & 0.429 & \\
$C_2$ & 0.063 & 0.011 & $-0.017$ & $-0.032$ & $-0.041$ & $-0.048$ & $-0.056$ & $-0.064$ & $-0.077$ & \\
$C_3$ & $-0.002$ & $-0.024$ & $-0.033$ & $-0.039$ & $-0.042$ & $-0.047$ & $-0.051$ & $-0.056$ & $-0.063$ & \\
$C_4$ & 0.008 & $-0.029$ & $-0.049$ & $-0.060$ & $-0.067$ & $-0.073$ & $-0.078$ & $-0.086$ & $-0.094$ & \\
$C_5$ & 0.997 & 0.939 & 0.901 & 0.875 & 0.856 & 0.841 & 0.829 & 0.819 & 0.814 & \\
$C_6$ & 0.024 & 0.015 & 0.007 & 0.002 & $-0.002$ & $-0.004$ & $-0.005$ & $-0.008$ & $-0.005$ & \\
$C_7$ & $-0.288$ & $-0.267$ & $-0.255$ & $-0.248$ & $-0.245$ & $-0.245$ & $-0.247$ & $-0.250$ & $-0.260$ & \\
\midrule
$g_{{^1S_0}}$     & 0.086  & $-0.031$ & $-0.058$ & $-0.051$ & $-0.036$ & $-0.020$ & $-0.010$ & $-0.011$& $-0.032$ & \\
$g_{{^1S_0}}^\prime$ & $-1.173$ & $-0.532$ & $-0.262$ & $-0.141$ & $-0.086$ & $-0.063$ & $-0.057$ & $-0.062$ & $-0.077$ & \\
$g_{{^3S_1}}$     & 0.064  & 0.115    & $-0.053$ & $-0.090$ & $-0.078$ & $-0.051$ & $-0.024$ & $-0.012$& $-0.041$& \\
$g_{{^3S_1}}^\prime$ & $-2.416$ & $-1.056$ & $-0.487$ & $-0.235$ & $-0.120$ & $-0.069$ & $-0.054$ & $-0.059$ & $-0.087$& \\
$c_E$             & 1.063  & 0.436    & 0.176    & 0.081    & 0.069    & 0.097    & 0.167    & 0.180 & 0.222 & \\
\midrule
$E(^{3}$H)$^\mathrm{w/o}_\mathrm{3NF}$ & $-7.886$ & $-8.127$ & $-8.283$ & $-8.357$ & $-8.350$ & $-8.277$ & $-8.163$ & $-7.961$ & $-7.776$ & \\
$E(^{3}$H)        & $-8.482$ & $-8.482$ & $-8.482$ & $-8.482$ & $-8.482$ & $-8.482$ & $-8.482$ &$-8.482$ &$-8.482$ & $-8.482$ \\
\midrule
$E(^{4}$He)$^\mathrm{w/o}_\mathrm{3NF}$       & $-24.0(1)$ & $-25.8(1)$ & $-26.9(1)$ & $-27.5(1)$ & $-27.6(1)$ & $-27.1(2)$ & $-26.8(2)$ & 
$-26.0(4)$&
$-25.6(6)$&  \\
$E(^{4}$He)       & $-26.9(1)$ & $-27.6(1)$ & $-28.0(1)$ & $-28.1(1)$ & $-28.2(1)$ & $-28.2(2)$ & $-28.9(2)$ & 
$-27.9(5)$&
$-27.6(6)$& $-28.3$ \\
\midrule
$E(^{16}$O)$^\mathrm{w/o}_\mathrm{3NF}$ & $-96.9(11)$ & $-105.1(10)$ & $-114.4(9)$ & $-121.6(9)$ & $-123.7(9)$ & $-121.5(9)$ & $-116.6(10)$ & $-114.6(15)$ & $-112.2(30)$ & \\
$E(^{16}$O)       & $-118.6(11)$ & $-122.0(10)$ & $-126.2(10)$ & $-129.4(10)$ & $-130.0(10)$ & $-129.5(10)$ & $-129.5(10)$ & $-128.9(17)$ & $-130.2(31)$ & $-127.6$ \\
\bottomrule	 
\end{tabular}
}
     \label{tab:GIRQ4LECs}
    \end{table*}

Recent calculations using the same interactions as in the present work showed that GIR terms substantially improve the renormalization group (RG) invariance of the $^4$He binding energy~\cite{Shi2026_PLB874-140303}. Without GIR terms, $^4$He was overbound by approximately $1$~MeV, with the discrepancy scaling as $\mathcal{O}(\Lambda^{-2})$. Including the leading-order GIR terms removed this discrepancy. These findings motivate an examination of whether the cutoff stability and agreement with experiment reported in the main text persist after explicitly restoring Galilean invariance.

\begin{table*}[!t]
\centering
\caption{Expectation values of selected operators in the Wigner-SU(4) ground states. The interactions are regularized with $\Lambda = 400$~MeV. The column \(E_\mathrm{exp}\) lists the experimental total energies (negative of the binding energies) in MeV. The ratios retain their signs. The OPE three-body contribution
and its ratio to $E_{\rm 3N}$ are both displayed with a factor of 200.
Experimental energies are taken from AME2020~\cite{Wang2021_CPC45-030003}.}
\label{tab:universality}
\small
\setlength{\tabcolsep}{3.4pt}
\resizebox{\linewidth}{!}{%
\begin{tabular}{l c c c c c c c c c c c c c c c}
\toprule
\rule[-2ex]{0pt}{6ex} & $E_\mathrm{SU4}$ & $E_\mathrm{exp}$ & $E_{Q^2}$ & $E_{Q^4}$ & $\Delta E_{Q^4}$ & $K_2$ & $K_\infty$ & $\Delta K_\infty$ & $E_\mathrm{GIR}$ & $200E_\mathrm{3N}^\mathrm{OPE}$ & $E_\mathrm{3N}$ & $\dfrac{\Delta E_{Q^4}}{E_\mathrm{3N}}$ & $\dfrac{\Delta K_\infty}{E_\mathrm{3N}}$ & $\dfrac{E_\mathrm{GIR}}{E_\mathrm{3N}}$ & $\dfrac{200E_\mathrm{3N}^\mathrm{OPE}}{E_\mathrm{3N}}$ \\
\midrule
\(^{4}\mathrm{He}\)  & $-23.68$ & $-28.32$ & $-52.27$ & $-46.15$ & $6.12$  & $39.66$ & $39.94$ & $0.273$ & $-1.51$  & $1.10$   & $-9.18$  & $-0.667$ & $-0.030$ & $0.165$ & $-0.119$ \\
\(^{16}\mathrm{O}\)  & $-103.2$ & $-127.6$ & $-294.2$ & $-262.5$ & $31.7$  & $233.5$ & $235.1$ & $1.634$ & $-11.9$  & $5.82$   & $-51.0$  & $-0.622$ & $-0.032$ & $0.233$ & $-0.114$ \\
\(^{24}\mathrm{Mg}\) & $-179.2$ & $-198.3$ & $-431.1$ & $-385.2$ & $45.9$  & $325.8$ & $327.9$ & $2.176$ & $-15.8$  & $7.92$   & $-69.5$  & $-0.660$ & $-0.031$ & $0.227$ & $-0.114$ \\
\(^{32}\mathrm{Mg}\) & $-210.5$ & $-249.7$ & $-585.3$ & $-527.6$ & $57.7$  & $467.8$ & $470.9$ & $3.061$ & $-22.2$  & $10.18$  & $-89.9$  & $-0.640$ & $-0.034$ & $0.247$ & $-0.113$ \\
\(^{40}\mathrm{Ca}\) & $-336.4$ & $-342.1$ & $-824.2$ & $-744.6$ & $79.6$  & $610.5$ & $614.6$ & $4.027$ & $-32.6$  & $14.48$  & $-135.9$ & $-0.586$ & $-0.030$ & $0.240$ & $-0.107$ \\
\bottomrule
\end{tabular}
}
\end{table*}

\begin{figure*}
        \centering  
        \includegraphics[width=0.85\textwidth]  {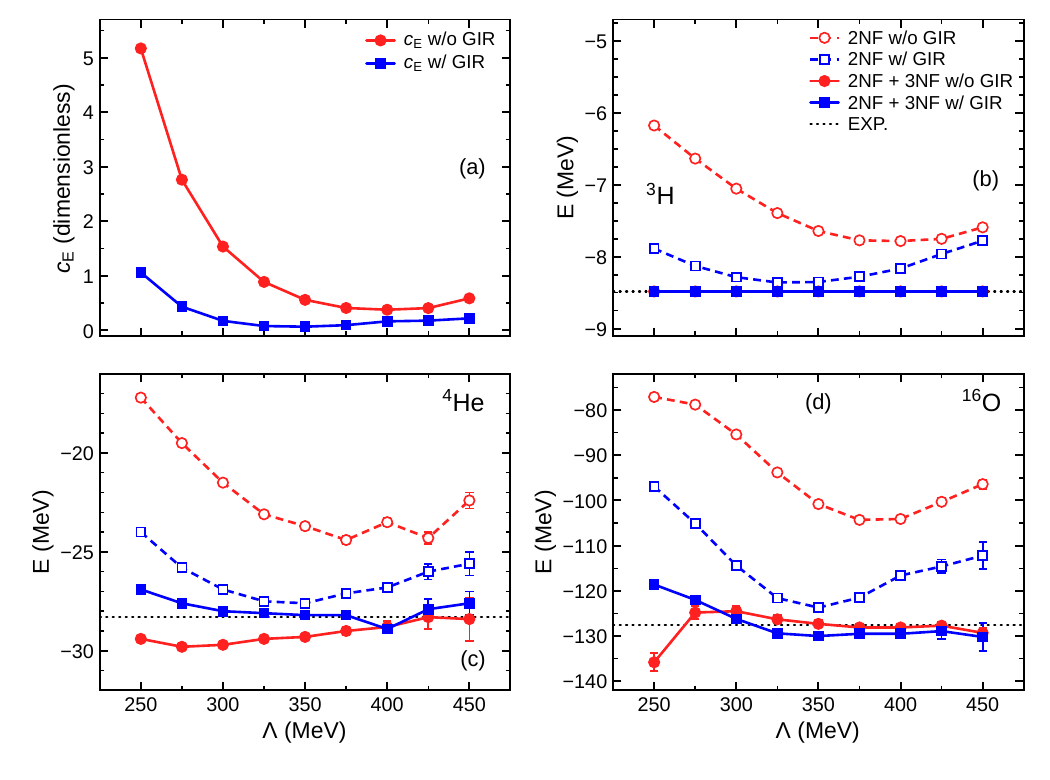}
        \caption{(a) Fitted three-body LEC $c_E$ with and without the Galilean invariance restoration terms.
        (b-d) Binding energies of $^3$H, $^4$He and $^{16}$O calculated with two-body force only (open symbols) and two- plus three-body forces (full symbols). Circles and squares denote the results without and with GIR terms, respectively. The dotted lines indicate the experimental values. }
        \label{fig:GIRQ4}
    \end{figure*}

For the actual $p^6$ regulator, we use the leading GIR structures
of Ref.~\cite{Liu2025_EPJA61-85},
\begin{equation}\label{eq:leading-GIR}
V_{\rm GIR}=P_{\rm cm}^2
[g_1+g_2(\bm\sigma_1\cdot\bm\sigma_2)]f_{\rm abs}.
\end{equation}
The Gaussian factors above serve only to illustrate the origin of
the boost dependence. Equation~\eqref{eq:leading-GIR} instead uses
the same regulator as the actual contact terms. The S-wave
coefficients are $g_{{}^1S_0}=g_1-3g_2$ and
$g_{{}^3S_1}=g_1+g_2$. They are fixed by reducing the dependence
of the S-wave scattering parameters on $P_{\rm cm}$.
Since these operators vanish at $P_{\rm cm}=0$, the original
center-of-mass fit is unchanged. Their implementation by FFT
follows Ref.~\cite{Liu2025_EPJA61-85}.

Fig.~\ref{fig:SwaveGIR} compares the NN $S$-wave phase shifts in the center-of-mass frame ($P_\mathrm{cm}=0$) and a moving frame ($P_\mathrm{cm}=150$~MeV), with and without GIR corrections. 
We fix $\Lambda = \Lambda_\pi = 300$~MeV, for which the Galilean breaking effects are relatively large.
Including the GIR terms restores agreement between the two frames. The $S$ waves are particularly sensitive to Galilean-invariance breaking because shallow bound states, such as the deuteron, and near-threshold virtual states require an all-order resummation of repeated scattering. The associated bubble resummation can amplify sensitivity to regulator-induced changes, requiring comparatively large GIR corrections.
In higher partial waves, such as $P$ and $D$ waves, the contact interactions considered here are comparatively perturbative, leading to weaker regulator dependence and smaller boost effects. Fig.~\ref{fig:PwaveGIR} illustrates this behavior for the $^{3}P_0$ channel. The phase shifts at $P_\mathrm{cm}=150$~MeV deviate only slightly from the center-of-mass results, with visible differences appearing only at relatively large momenta near $P_\mathrm{rel}\approx200$~MeV. By contrast, differences in the $S$ waves are already apparent near threshold.

These comparisons motivate retaining GIR corrections only in the $S$ waves in the present work. For simplicity, we neglect GIR corrections to $S$--$D$ mixing. The GIR coefficients are first fitted in calculations with OPEP
omitted. The local OPEP itself is Galilean invariant, but its iteration
with the contact terms can affect the residual boost dependence.
Figure~\ref{fig:SwaveGIR} therefore provides a separate check with
OPEP included. A complete GIR treatment, including higher partial waves and mixing terms, is left to future work.

The corresponding momentum dependence of the $^{1}S_0$ and $^{3}S_1$ scattering lengths is shown in Fig.~\ref{fig:011}. This diagnostic omits both OPEP and $S$--$D$ mixing while retaining
the original contact LECs. Its center-of-mass scattering lengths are
therefore not required to reproduce the experimental values.

To extend the restoration to the larger nucleon momenta relevant for medium-mass nuclei such as $^{16}$O, we supplement the leading GIR terms with their next momentum-dependent
corrections. The corresponding S-wave kernels are
\begin{align*}
V_{\mathrm{GIR}}^{1\mathrm{S}0} &=P_{\rm cm}^2
[g_{1\mathrm{S}0} + g_{1\mathrm{S}0}^\prime (p^2 + p^{\prime2})] f_{\rm abs}(p_1',p_1, p_2', p_2), 
\\
V_{\mathrm{GIR}}^{3\mathrm{S}1} &= P_{\rm cm}^2
[g_{3\mathrm{S}1} + g_{3\mathrm{S}1}^\prime (p^2 + p^{\prime2})] f_{\rm abs}(p_1',p_1, p_2', p_2),
\end{align*}
where $f_{\rm abs}(p_1', p_1, p_2',p_2)$ is the same absolute-momentum regulator used elsewhere in this work. The coefficients $g_{^1S_0/^3S_1}$ and $g_{^1S_0/^3S_1}'$ are determined by requiring the $S$-wave neutron-proton phase shifts to remain independent of $\bm{P}_{\rm cm}$ over the momentum range relevant to these nuclei. At each cutoff, we refit the three-body LEC $c_E$ to reproduce the experimental triton binding energy. The resulting running LECs are summarized in Tab.~\ref{tab:GIRQ4LECs}.
Since the GIR terms vanish in the two-body center-of-mass frame, the original two-body contact LECs remain identical to those used in the main text and shown in Tab.~\ref{tab:LECsEnergies}. 
After the inclusion of GIR terms, the required three-body forces are significantly reduced.

\begin{figure*}
    \centering  
    \includegraphics[width=0.85\textwidth]  {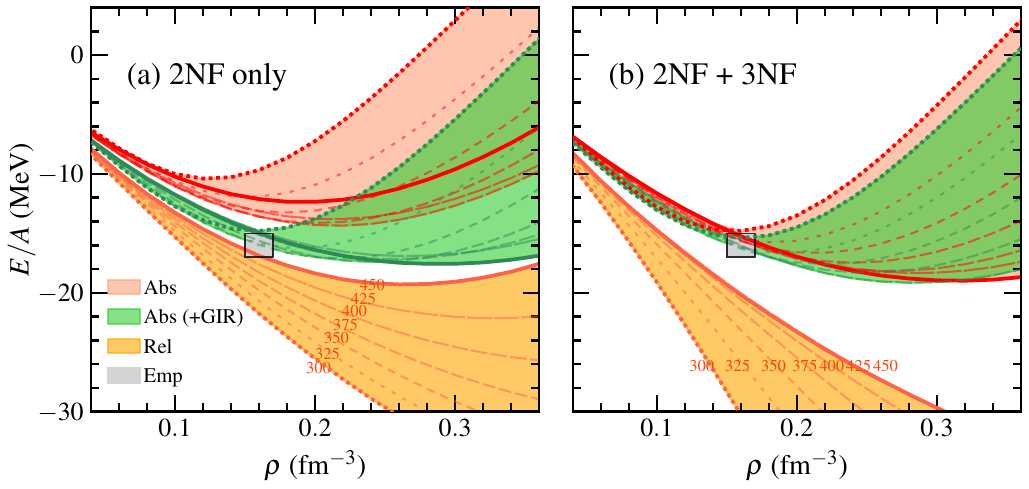}
    \caption{The equations of state of symmetric 
    nuclear matter calculated with the chiral nuclear 
    forces with absolute-momentum (abs) regulator, 
    abs regulator plus Galilean invariance restoration
    (GIR) terms and relative-momentum (rel) regulator. 
    Panels (a) and (b) show the results obtained with 2NF only 
    and 2NF + 3NF respectively.}
        \label{fig:GIRQ4SNM}
\end{figure*}

The binding energies of $^3$H, $^4$He, and $^{16}$O are shown in Tab.~\ref{tab:GIRQ4LECs} and Fig.~\ref{fig:GIRQ4}. The GIR interactions are implemented using recently developed techniques \cite{Liu2025_EPJA61-85}. 
Although the GIR terms induce non-negligible effects in these results, refitting $c_E$ to the triton energy largely compensates the
GIR-induced energy shifts, as expected from the RG flow analysis.
With both GIR terms and three-body forces included, the energies exhibit only modest cutoff dependence over $300$~MeV $\le \Lambda \le 450$~MeV and remain close to the experimental values. The residual discrepancy for $^4$He is attributed mainly to charge-symmetry-breaking effects, which are omitted here for simplicity. In Ref.~\cite{Shi2026_PLB874-140303}, including these effects yielded a $^4$He binding energy within $100$~keV of the experimental value. Explicit restoration of Galilean invariance therefore preserves the cutoff stability and agreement with experiment reported in the main text. In Fig.~\ref{fig:GIRQ4SNM} we present the corresponding results for SNM. The conclusions drawn in the main text do not change after inclusion of the GIR corrections.

\section{Higher-order corrections}

A natural question is whether the observed cutoff stability persists upon including higher-order corrections. Although a complete high-precision calculation incorporating all such terms is challenging, perturbative estimates can provide insight into their effects. Here, we examine five representative contributions.

First, the lattice kinetic energy can be implemented with different levels of improvement, ranging from nearest-neighbor finite differences to a fast Fourier transform (FFT). We denote by $K_2$ the kinetic-energy operator constructed with a five-point stencil along each spatial direction, and by $K_\infty$ its FFT counterpart. Their difference, $\Delta K_\infty = K_\infty - K_2$, probes high-momentum corrections absent from $K_2$. Such improvements to the lattice kinetic energy has been used to illustrate the Symanzik improvement program~\cite{Gattringer2010_LNP788}.

Second, we refit the interaction after including the $15$ additional $\mathcal{O}(Q^4)$ contact terms described in Ref.~\cite{Liu2025_EPJA61-85}. We then evaluate the change in the contact-interaction contribution to the energy between the $\mathcal{O}(Q^2)$ and $\mathcal{O}(Q^4)$ fits, denoted by $\Delta E_{Q^4} = E_{Q^4} - E_{Q^2}$.

Finally, we consider the GIR terms introduced in the previous section and the two three-nucleon interactions proportional to $c_D$ and $c_E$. The energy contributions of the latter two terms are denoted by $E_\mathrm{3N}^\mathrm{OPE}$ and $E_\mathrm{3N}$, respectively.

We consider five representative nuclei, $^4$He, $^{16}$O, $^{24}$Mg, $^{32}$Mg, and $^{40}$Ca. The first-order energy corrections are evaluated as expectation values of the corresponding operators in the ground states of the Wigner-SU(4) Hamiltonian. All operator coefficients are fixed in advance and kept unchanged across the five nuclei. The results are listed in Tab.~\ref{tab:universality}, with $E_\mathrm{3N}^\mathrm{OPE}$ multiplied by $200$ for presentation.

To assess the universality of these contributions, we examine their ratios to $E_\mathrm{3N}$. Apart from $E_\mathrm{GIR}/E_\mathrm{3N}$ in $^4$He, the ratios remain nearly constant from light nuclei to medium-mass nuclei up to $^{40}$Ca, revealing a remarkable degree of universality across this mass range.

\begin{figure}
        \centering  
        \includegraphics[width=1.0\columnwidth]  {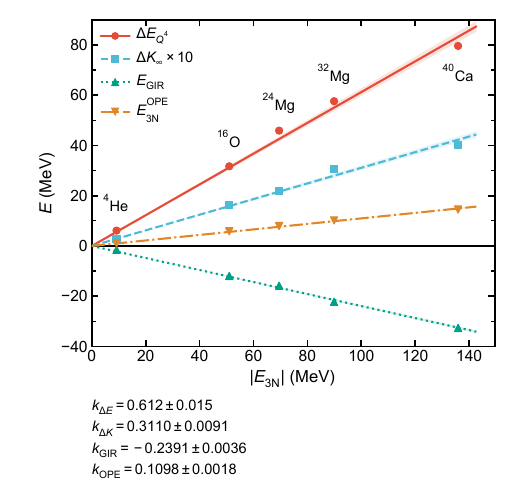}
        \caption{Correlations between various energy corrections and the three-body forces. $\Delta K_\infty$ is multiplied by 10 and
        $E_{\rm 3N}^{\rm OPE}$ by 200 for visibility.
        Fits are constrained to pass through the origin.
        The quoted slope uncertainties are standard errors of the
        unweighted fits.}
        \label{fig:UniversalityCorrelation}
    \end{figure}

\begin{figure}
        \centering  
        \includegraphics[width=1.0\columnwidth]  {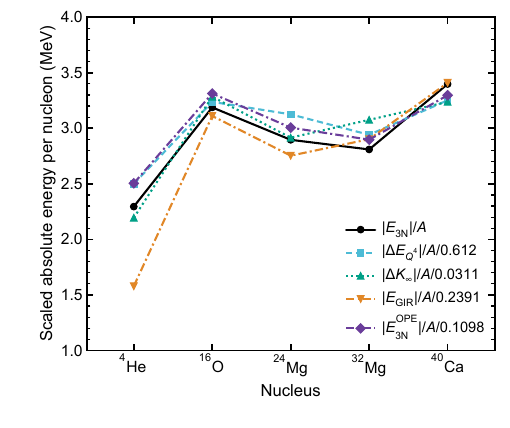}
        \caption{Universality of first-order perturbative energy corrections from selected higher-order operators. Each contribution is rescaled by an overall factor determined from the corresponding slope in Fig.~\ref{fig:UniversalityCorrelation}.}
        \label{fig:rescaledquantities}
    \end{figure}

Fig.~\ref{fig:UniversalityCorrelation} shows these contributions as functions of $|E_\mathrm{3N}|$, revealing approximately linear correlations across the nuclei considered. Fig.~\ref{fig:rescaledquantities} presents the corresponding quantities rescaled using the slopes extracted from these correlations. The rescaled contributions follow closely similar trends, suggesting that the higher-order energy corrections are strongly correlated across different nuclei. At the level of these first-order binding-energy corrections, their effects may therefore be approximately absorbed into adjustments of one or a few three-body couplings.

The observed universality admits a natural interpretation in terms of Wilsonian RG flow. As momentum modes between the cutoff and the characteristic scale of the system are integrated out, low-energy observables become progressively less sensitive to irrelevant directions in coupling space. Their dominant dependence can then be described by a low-dimensional manifold parametrized by a small number of leading LECs. This does not require higher-order operators to disappear under the flow. Along a renormalized trajectory, their couplings can remain nonzero while becoming correlated with the leading couplings, with independently variable components suppressed at the accuracy considered.

Adding higher-order operators and refitting to the same low-momentum observables can therefore yield approximately equivalent low-energy descriptions, provided that the accompanying many-body interactions are treated consistently. For example, the $\mathcal{O}(Q^2)$ and $\mathcal{O}(Q^4)$ contact interactions considered here are calibrated to the same NN phase shifts. The quantity of interest is the residual difference between the refitted Hamiltonians, whose first-order energy contribution is $\Delta E_{Q^4}$, rather than the full contact contribution $E_{Q^4}$. Our numerical results suggest that the dominant effects of this residual difference on the binding energies considered here can be approximately absorbed into adjustments of a small number of three-body couplings. The GIR corrections discussed in the main text provide another example of such compensation.

These observations motivate the conjecture that only a few independent combinations of three- and higher-body couplings control the energy shifts examined here. They also raise the question of how much independent low-energy information is introduced by additional three-body operators used to recalibrate nuclear EFT to nuclear-structure observables. After coarse-graining, some of these operators may produce nearly indistinguishable effects on the selected observables, limiting how independently their coefficients can be determined. Further universality studies based on high-precision \textit{ab initio} calculations across a broader range of nuclei and observables are needed to establish the extent of this behavior.

\end{document}